%% file: main.tex
\documentclass[runningheads]{llncs}

\usepackage[mobile]{eccv}

\usepackage{eccvabbrv}

\usepackage{graphicx}
\usepackage{booktabs}

\usepackage[accsupp]{axessibility}  
\input{preamble}

\usepackage{hyperref}

\usepackage{orcidlink}

\begin{document}

\title{Pairwise Distance Distillation for Unsupervised Real-World Image Super-Resolution} 

\titlerunning{Pairewise Distance Distillation}

\author{Yuehan Zhang\inst{1}\orcidlink{0000-0002-5017-0097} \and
Seungjun Lee\inst{2}\orcidlink{0009-0009-7753-1962} \and
Angela Yao\inst{1}\orcidlink{0000-0001-7418-6141}}

\authorrunning{Y.~Zhang et al.}

\institute{National University of Singapore\\ \email{\{zyuehan,ayao\}@comp.nus.edu.sg}\and
Korea University\\
\email{9penguin9@korea.ac.kr}}

\maketitle
\begin{abstract}
Standard single-image super-resolution creates paired training data from high-resolution images through fixed downsampling kernels. However, real-world super-resolution (RWSR) faces unknown degradations in the low-resolution inputs, all the while lacking paired training data. Existing methods approach this problem by learning blind general models through complex synthetic augmentations on training inputs; they sacrifice the performance on specific degradation for broader generalization to many possible ones. We address the unsupervised RWSR for a targeted real-world degradation. We study from a distillation perspective and introduce a novel pairwise distance distillation framework.
Through our framework, a model specialized in synthetic degradation adapts to target real-world degradations by distilling intra- and inter-model distances across the specialized model and an auxiliary generalized model. Experiments on diverse datasets demonstrate that our method significantly enhances fidelity and perceptual quality, surpassing state-of-the-art approaches in RWSR. The source code is available at \url{https://github.com/Yuehan717/PDD}.

\end{abstract}

\input{sec_TS/1_intro}

\input{sec_TS/2_related_works}

\input{sec_TS/3_method}

\input{sec_TS/4_experiments}

\section{Conclusion}
This paper studies unsupervised real-world image super-resolution from the distillation perspective. The approach considers both the real-world knowledge of a generalized SR model and the low-level characteristics of synthetic predictions of the specialist model, adapting a synthetic domain specialist to target real-world degradation. 
The core of our method is the Pairwise Distance Distillation. By enforcing the consistencies for intra- and inter-model distances, our method aims at pushing the low-level characteristics of the specialist's real-world predictions towards its synthetic predictions. 
As a learning scheme, our method can improve off-the-shelf models regarding fidelity and perception on multiple real-world datasets. 
Additionally, we emphasize our approach to offering a new perspective toward addressing real-world super-resolution challenges.

\subsubsection{Acknowledgement} This research/project is supported by the Ministry of Education, Singapore, under the Academic Research Fund Tier 1 (FY2022).
%
%
\bibliographystyle{splncs04}
\bibliography{main}
\end{document}

%% file: preamble.tex
\usepackage{graphicx}
\usepackage{amsmath}
\usepackage{amssymb}
\usepackage{makecell}
\usepackage{subcaption}
\usepackage{caption}
\usepackage{multirow}
\usepackage{multicol}
\usepackage[dvipsnames]{xcolor}
\usepackage{soul}
\usepackage{comment}
\usepackage{bbding}
\usepackage{makecell}
\usepackage{xspace}
\usepackage{soul}

\makeatletter
\DeclareRobustCommand\onedot{\futurelet\@let@token\@onedot}
\def\@onedot{\ifx\@let@token.\else.\null\fi\xspace}

\def\eg{\emph{e.g}\onedot} 
\def\ie{\emph{i.e}\onedot}

\makeatother
\usepackage[normalem]{ulem}

\newcommand{\Size}[3]{\mathbb{R}^{#1\times\!#2\times\!#3}}

\newcommand{\Pred}[2]{\hat{Y}_{#1}^{#2}}
\newcommand{\Dct}[1]{d^{ij}_{#1}}
\newcommand{\Dap}[1]{\Delta^{ij}_{#1}}
\newcommand{\Feat}[1]{\Phi_{ij}(#1)}
\newcommand{\Frobenius}[1]{\|#1\|_\mathcal{F}}
\newcommand{\GT}{{Y}^{L}}
\newcommand{\wvt}[1]{W(#1)_i}

\newcommand{\drop}[2]{$#1_{\textcolor{blue}{(#2)}}$}
\newcommand{\improve}[2]{$#1_{\textcolor{red}{(#2)}}$}
\newcommand{\Best}[1]{\textbf{#1}}
\newcommand{\Second}[1]{\underline{#1}}
\newcommand{\InsertSubfig}[2]{ \begin{subfigure}[t]{#1\linewidth}
    \centering
        \includegraphics[width=\textwidth]{#2}
    \end{subfigure}}
\newcommand{\InsertSubfigWithCap}[3]{ \begin{subfigure}[t]{#1\linewidth}
    \centering
        \includegraphics[width=\textwidth]{#2}
        \subcaption{#3}
    \end{subfigure}}
\newcommand{\InsertSubfigWithCapWithLabel}[4]{ \begin{subfigure}[t]{#1\linewidth}
    \centering
        \includegraphics[width=\textwidth]{#2}
        \subcaption{#3}
        \label{#4}
    \end{subfigure}}


%% file: sec_TS/1_intro.tex
\section{Introduction}
\label{sec:intro}

Single-image super-resolution (SISR) predicts high-resolution (HR) images from low-resolution (LR) counterparts. The standard SISR model addresses predefined downsampling kernels, \eg bicubic interpolation. However, real-world scenarios of SISR (RWSR) encompass unknown degradations in LR images, including blur, noise, and JPEG compression artifacts~\cite{wang2021real,zhang2023crafting,zhang2021designing} with diverse combinations. The various and unknown degradations pose additional challenges when learning an RWSR model.

\begin{figure}[t]
\centering
\includegraphics[width=0.85\linewidth]{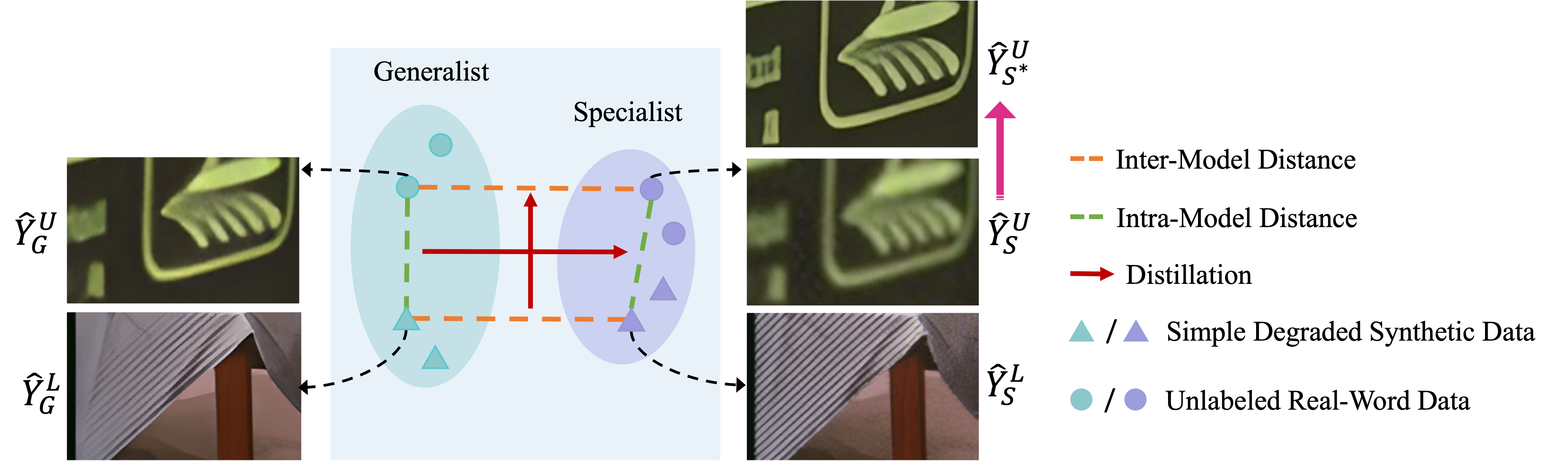}
\caption{$\Pred{G}{U}$ and $\Pred{G}{L}$ are reconstructions for real-world and bicubic-interpolated (BI) inputs using a blind generalized model (generalist); $\Pred{S}{U}$ and $\Pred{S}{L}$ are counterparts using a standard SR model (bicubic specialist). The specialist enhances the BI inputs with clearer details, while the generalist does better for the real-world one. We distill the intra- and inter-model distances for an improved real-world reconstruction $\Pred{S^*}{U}$.}
\label{fig:teaser}
\end{figure}

Existing RWSR methods focus on blind generalization. Such methods synthesize paired training data with extensive and harsh degradations, involving multiple rounds of random blurring, added noise, resizing, and compression~\cite{wang2021real,zhang2021designing}.
The complex degradation pipeline allows the model to generalize to diverse unknown conditions. We refer to such a model addressing various degradations as a \textit{generalist}. However, as much of the model's capacity is devoted to handling multiple conditions, these ``jack-of-all-trades'' models come with performance trade-offs~\cite{zhang2023crafting}, \ie, they have inferior performance compared to the model only optimized for the tested degradation ($\Pred{G}{L}$ and $\Pred{S}{L}$ in \cref{fig:teaser}). 

SR models can be optimized for specific and known input degradations through supervised learning~\cite{wang2018esrgan,liang2021swinir,gu2019blind,huang2020unfolding}. The LR-HR pairs for training are created by degrading HR images with fixed kernels, \eg, bicubic interpolation, and Gaussian blur. We refer to models adept at the specific input degradation as a \textit{specialist}.  
Degradations in RWSR datasets~\cite{cai2019toward,wei2020component} always belong to a specific domain~\cite{zhang2023crafting}. For example, images captured by a specific camera model tend to feature consistent sensor noise~\cite{cai2019toward}. However, the formulation of a real-world degradation is always unknown, posing difficulties in creating paired training data. As such, learning specialist models for such real-world domains is challenging, and strategies that do not require direct supervision are crucial.

This paper tackles unsupervised RWSR from a knowledge distillation perspective. Distillation has been successful for domain adaptation in high-level tasks~\cite{gao2022cross,wang2020go,liu2020adaptive}, but it is under-explored in low-level vision. Our distillation framework adapts a specialist model of known synthetic degradation to target real-world degradations. We use an auxiliary generalist model to provide generalization knowledge of real-world degradations that the specialist lacks.
However, naive distillation only transfers knowledge from the generalist model; we aim to improve beyond the generalist's moderate performance. To that end, we also consider the low-level characteristics of the specialist's predictions on synthetic samples as a reference for the distillation.

Our method uses the strength of both the specialist and the generalist by exploring the relationships between their predictions.
Given a generalist and a specialist for synthetic degradation, we consider predictions of synthetic samples and real-world samples from both models (see \cref{fig:teaser}). The generalist model has smaller gaps in qualities of real-world and synthetic predictions, featuring close low-level characteristics between domains~\cite{liu2021discovering,liu2023evaluating}.
In contrast, the specialist outputs distanced low-level characteristics but high-quality predictions for synthetic samples ($\Pred{S}{L}$ in \cref{fig:teaser}), providing valuable low-level characteristics for reference.

We propose to let the specialist imitate prediction \textit{relationships} from the generalist - that is, push the specialist's real-world predictions ($\Pred{S}{U}$ in \cref{fig:teaser}) to have low-level characteristics similar to its high-quality synthetic predictions.
If such similarity is satisfied by the specialist, it would exhibit two consistencies. First, for two input samples, the difference between their predictions from the same model is intra-model distance; such distances would be consistent for the specialist and the generalist. Secondly, for a given input sample, the difference between generalist and specialist predictions is inter-model distance; such distances would be consistent whether the sample is synthetic or real-world.
These intra- and inter-model distance consistencies form the basis of our pairwise distance distillation framework. Together with designs on configurations of specialist and generalist models, our adapted specialist model experimentally shows significant improvements over the generalist model.

\noindent\textbf{Contributions.} To summarize, the contributions of our work are three-fold:
\begin{itemize}
    \item To the best of our knowledge, we are the first to propose a distillation perspective for unsupervised RWSR to combine generalist and specialist models.
    \item We propose a novel pairwise distance distillation framework, emphasizing transferring intra-model and inter-model distances to enhance the specialist's performance in real-world scenarios.
    \item Experiments on three benchmarks demonstrate that our approach improves the off-the-shelf models regarding fidelity and perceptual performances.
\end{itemize}

%% file: sec_TS/2_related_works.tex
\section{Related Works}
\label{sec:related_works}

\noindent\textbf{Single-Image Super-Resolution (SISR)} methods use deep neural networks of various architectures, such as residual networks~\cite{lim2017enhanced,wang2018esrgan} and transformers~\cite{liang2021swinir, wang2023omni}. Despite improving the standard SISR performance with dedicated architectures, these methods struggle to generalize to real-world scenarios~\cite{wei2021unsupervised,wang2021unsupervised}.

\noindent\textbf{Real-World Image Super-Resolution (RWSR)} aims to address unknown input degradations. The infeasibility of creating paired training data without known kernels presents a significant challenge in learning specialized models for RWSR.
There are two primary strategies in existing RWSR. The first
assumes that networks trained on diverse and challenging synthetic degradations will effectively generalize to real-world data~\cite{wang2021real,zhang2021designing}. 
Even though real-world degradations are complex, within a specific setting or deployment, they tend to be limited in the domain and less diverse than the training pipelines. 
Consequently,~\cite{zhang2023crafting} proposes customizing the synthetic pipeline to better match specific real-world data. Nonetheless, an unresolved gap exists between synthesized and real-world degradations. Instead, our method operates directly on real-world data and avoids the intrinsic gap from synthetic degradation.

The second strategy connects real-world and synthetic degradation with transfer learning. Most works use image-to-image translation~\cite{wei2021unsupervised, romero2022unpaired,fritsche2019frequency} to adapt the synthetic degradation to real-world ones. Such methods resort to intricate designs like CycleGAN~\cite{zhu2017unpaired}, which struggle to replicate real-world degradations~\cite{wang2021unsupervised} reliably.
Our proposed method explores feature distances between model outputs and bypasses the need to mimic real-world degradations. 

\noindent\textbf{Knowledge Distillation} is initially introduced for model compression~\cite{hinton2015distilling}. In low-level vision, several works~\cite{gao2018image,zhang2021data,park2021local} distill large models into more efficient ones by enforcing similarity between their internal features or predictions. 
Recently, knowledge distillation has served diverse purposes in high-level tasks like segmentation, including domain adaptation~\cite{nguyen2021unsupervised,li2023progressive,gao2022cross} and transfer learning~\cite{wang2020go,yang2022cross}. 
However, such applications are less discussed in low-level vision.

%% file: sec_TS/3_method.tex
\section{Method}
\label{Sec:Method}
Our method adapts the specialist in synthetic degradation to an unlabeled real-world domain. We use the low-level characteristics of synthetic predictions from the specialists and the knowledge of the generalist model.
\cref{fig:overview} is the overview of our approach and the novel Pairwise Distance Distillation (PDD) method, which enforces the consistency of the intra- and inter-model distances to improve the specialist ($M_S$)'s real-world predictions.
\cref{Subsec:preliminary} gives notions used in this section and \cref{Subsec:formulation} illustrates the unsupervised formulation. \cref{Subsec:distillation} explains the details of PDD and \cref{Subsec:full_method} provides the full optimization aims and discusses the static and Exponential Moving Average (EMA) versions of our method.

\subsection{Definitions \& Setup}
\label{Subsec:preliminary}
When provided with an LR input $X\in\Size{H}{W}{C}$ - where $H\times\!W$ is the spatial resolution and $C$ is the number of color channels - a SISR model aims to reconstruct an HR image $\hat{Y}\in\Size{rH}{rW}{C}$, scaling by a factor $r$. Due to the difficulty in obtaining paired training data, the conventional approach in SISR involves generating an LR image $X$ by downsampling the HR ground-truth $Y$, \ie, $X=D_{\downarrow} (Y)$. This downsampling operation $D_{\downarrow}$ typically refers to bicubic interpolation, although other alternatives exist~\cite{Agustsson_2017_CVPR_Workshops}.

In a more extensive setting, it is typical for the LR image to be affected by various factors, such as blurring, noise, compression artifacts, \textit{etc}. 
To exhibit these degradations, the LR image $X$ can be formed by applying $D_{\downarrow}$ along with one or multiple degradations $D_i$, in any order, \ie, 
\begin{equation}
X = D_1 \circ \dots \circ D_n (Y).
\end{equation}
Here, we include $D_{\downarrow}$ as one of the degradations $\{D_i\}_{i \in [1,n]}$ for simplicity. 
Now, let's consider two categories of degradation sets consisting of known degradations: $\mathcal{D}_S$, which comprises simple degradations with a few factors, and $\mathcal{D}_G$ characterized by complex degradations with a wide range of factors: 
\begin{equation}
\mathcal{D}_S = \{D_i\}_{i \in \mathcal{S}},\; 
\mathcal{D}_G = \{D_i\}_{i \in \mathcal{G}}, \quad \text{where } |\mathcal{G}| \gg |\mathcal{S}|.
\end{equation}

\noindent
In the simplest case for $\mathcal{D}_S$, the set $\mathcal{D}_\mathcal{S} = \{D_{\downarrow}\}$, which represents the standard SISR setup. For the broader case, $\mathcal{D}_G$ should encompass diverse and challenging degradations beyond $\{D_{\downarrow}\}$ to ensure that models trained on this data generalize well to blind settings~\cite{wang2021real,zhang2021designing}.

Consider two models, $M_S$ and $M_G$, trained on datasets constructed with $\mathcal{D}_S$ and $\mathcal{D}_G$ respectively. Both models are optimized with common supervised SR learning losses~\cite{wang2018esrgan,ledig2017photo} including MSE-based loss, VGG loss, and adversarial training (see \cref{eq:L_L}). 
Since $\mathcal{D}_S$ comprises a narrower range of degradations compared to $\mathcal{D}_G$, $M_S$ is a ``\textit{specialist}'' model for the limited domain defined by $\mathcal{D}_S$. On the other hand, $M_G$ is a ``\textit{generalist}'' model as it achieves moderate performance across the broader domain derived by $\mathcal{D}_G$. 
Comparing $M_S$ and $M_G$, we assume the specialist's superiority to the generalist in $\mathcal{D}_S$ and the generalist's better handling of unknown degradations. These assumptions are empirically validated in \cref{tab:assumption}.

\subsection{Unsupervised Learning Through Distillation}
\label{Subsec:formulation}
Our approach aligns with the unsupervised setup~\cite{wei2021unsupervised,romero2022unpaired,fritsche2019frequency}, wherein we work with a collection of real-world LR images alongside unpaired HR clean images.
We denote the set of LR images as $\{X^U\}$ with the same unknown degradations $\{D_i\}_{i\in\mathcal{U}}$. Additionally, we generate LR counterparts $\{X^L\}$ from HR images using specific degradations from $\mathcal{D}_S$, \eg bicubic interpolation.
The specialist $M_S$ is adept in restoring $X^L$, while $M_G$ performs moderately in restoring $X^L$ and $X^U$. 
We aim to adapt $M_S$ to $\{X^U\}$ with the aid of $M_G$.
As in~\cite{hinton2015distilling}, a naive distillation approach is having $M_S$ imitate the predictions from $M_G$ for unlabeled inputs: 
\begin{equation}
\begin{aligned}
\Pred{S}{L} &= M_S(X^L),\: \Pred{S}{U}= M_S(X^U),\: \Pred{G}{U} = M_G(X^U), \\
\mathcal{L}_{ND} &= \mathcal{L}_L(\Pred{S}{U},\Pred{G}{U}) + \lambda\mathcal{L}_L(\Pred{S}{L},Y^L),
\end{aligned}
\label{eq:baseline}
\end{equation}
where $Y^L$ represents the ground-truth image in the labeled domain, $\mathcal{L}_L$ denotes the supervised loss formulated in~\cref{eq:L_L}, and {$\lambda$ is a scale factor to balance between the distillation and primary objective of $M_S$}.
However, this simple imitation approach relies solely on information from the generalist $M_G$ and fails to harness the strengths of both models.

\begin{figure}[t]
    \centering
    \includegraphics[width=0.90\textwidth]{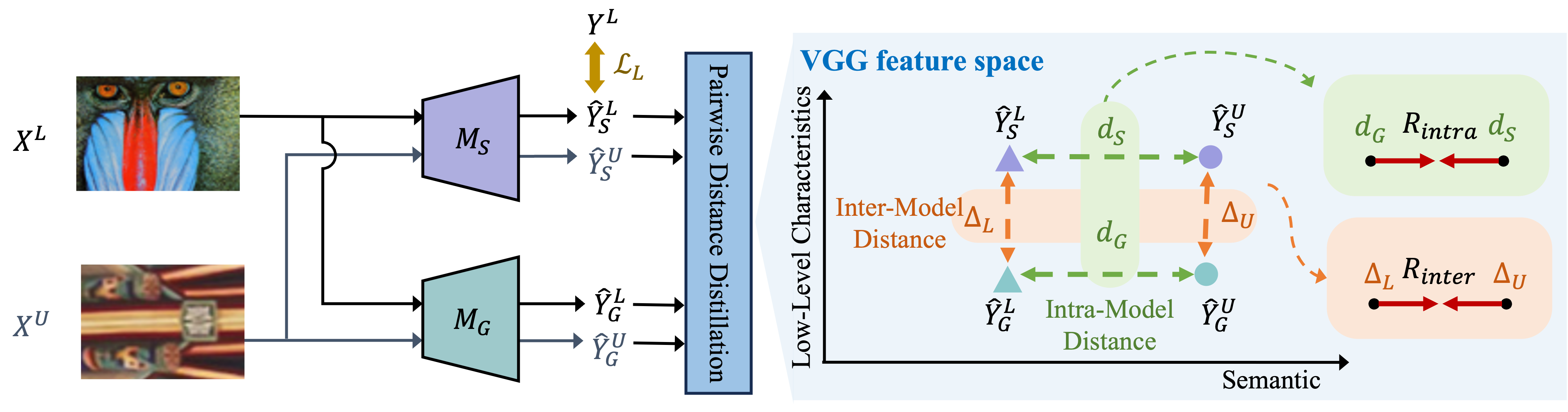}
    \caption{Schematic of Pairwise Distance Distillation (PDD). $X^U$ and $X^L$ are unlabeled (real-world) and labeled (synthetic) inputs. Model $M_G$ is a generalist trained with extensive synthetic pipeline, while $M_S$ specializes in $X^L$. The prediction $\Pred{S}{L}$ is supervised by its ground truth throughout training. PDD enforces the consistency between $\{\Delta_U,\:\Delta_L\}$ and between $\{d_S,\:d_G\}$ to improve $M_S$'s real-world performance.}
    \label{fig:overview}
\end{figure}

\subsection{Pairwise Distance Distillation (PDD)}
\label{Subsec:distillation}
Consider any pair of LR images - a real-world $X^U$ and a synthesized $X^L$. As shown in \cref{fig:overview}, we obtain four predictions by applying $M_S$ and $M_G$:
\begin{equation}
    \begin{aligned}
        \Pred{S}{U}&= M_S(X^U),\: \Pred{G}{U} = M_G(X^U),\\
        \text{and} \;\;
         \Pred{S}{L}&= M_S(X^L),\: \Pred{G}{L} = M_G(X^L).
    \end{aligned}
    \label{eq:predictions}
\end{equation}

{Our distillation explores the relationship between predictions in \cref{eq:predictions} to seek for knowledge combination. 
We use VGG features~\cite{simonyan2014very} of the predictions as our basis for exploration. VGG, as a classification model, inherently captures \textit{semantic} information. Yet it also captures \textit{low-level characteristics} related to image quality~\cite{wang2020deep}, \eg blur and sharpness, and is widely used in the perceptual loss for super-resolution  tasks~\cite{ledig2017photo,wang2018esrgan,wang2021real}.

As a generalist, $M_G$ outputs synthetic and real-world predictions with small quality gaps, which is characterized by the overlapping distributions of low-level characteristics of $\{\Pred{G}{L}\}$ and $\{\Pred{G}{U}\}$~\cite{liu2021discovering,liu2023evaluating}. Following~\cite{liu2021discovering}, \cref{subfig:dist_MG} projects the low-level features of $\{\Pred{G}{L}\}$ (DIV2K) and $\{\Pred{G}{U}\}$ (NTIRE20). The generalist's coverage strongly overlaps while the specialist's predictions are well-separated (see \cref{subfig:initial_MS}). The within-model relationship of the generalist exhibits the generalizability to real-world degradation.

On the other side, the high image quality of $\{\Pred{S}{L}\}$, due to $M_S$'s specialization, makes its low-level characteristics a valuable reference. To integrate the knowledge for real-world performance, we propose to make the specialist imitate the generalist's within-model relationship, that is, pushing the low-level characteristics of real-world predictions from $M_S$, $\{\Pred{S}{U}\}$, to be similar to that of its synthetic predictions $\{\Pred{S}{L}\}$.}
To achieve this aim, we first \textit{assume} such similarity is satisfied for $M_S$ and conclude the following consistencies about VGG feature distances between predictions:
\begin{enumerate}
\item Intra-model distances for predictions of the same input pair should be consistent across $M_S$ and $M_G$ (green shade in \cref{fig:overview}).
\item Inter-model distances for predictions of a single input should be consistent across synthetic and real-world domains (orange shade in \cref{fig:overview}). 
\end{enumerate}

Our rationales for these consistencies are as follows.  
First, we recall the low-level characteristics of predictions from the same model are similar. We discussed it for $M_G$ with \cref{subfig:dist_MG} and made such an assumption for $M_S$. Thus, the distance between the same model's predictions $\{\Pred{G}{L},\Pred{G}{U}\}$ (or $\{\Pred{S}{L},\Pred{S}{U}\}$), \ie intra-model distance, reflect mainly semantic differences due to their close low-level characteristics. As it naturally follows that the applied SR models should not change semantics, intra-model distances should be consistent for $M_S$ and $M_G$. 

Second, we note the low-level characteristics of predictions from the two models differ, \eg $\Pred{S}{L}$ has higher quality than the counterpart $\Pred{G}{L}$. The distance between the two predictions $\{\Pred{G}{L}, \Pred{S}{L}\}$ (or $\{\Pred{G}{U}, \Pred{S}{U}\}$) of a given sample, \ie inter-model distance, captures such differences in low-level characteristics, as semantics stay constant for the predictions of the same input. Given the similarity of low-level characteristics within a single model, the inter-model distances for synthetic and real-world samples should be consistent.

We structure the two consistencies as distillations on the intra- and inter-model distances. Our method encourages $M_S$'s real-world predictions $\{\Pred{S}{U}\}$ to have similar low-level characteristics as its synthetic predictions $\{\Pred{S}{L}\}$, taking it as a reference from high-quality images.

\noindent\textbf{Intra-model Distance Distillation} enforces consistency on the distances between predictions from the same model (green shading in \cref{fig:overview}). 
Consider the $\ell_1$ distance $\Dct{G}$ between 
predictions from the generalist $M_G$, \ie $\Pred{G}{L}$ and $\Pred{G}{U}$, 
in VGG feature space, and the distance $\Dct{S}$ for the specialist $M_S$ is similarly defined:
\begin{equation}
\begin{aligned}
    \Dct{G} &= \|\Feat{\Pred{G}{L}} - \Feat{\Pred{G}{U}}\|_1, \\
    \Dct{S} &= \|\Feat{\Pred{S}{L}} - \Feat{\Pred{S}{U}}\|_1,
\end{aligned}
\end{equation}
where $\Phi_{ij}$ refers to the $j$-th layer in the $i$-th residual block of VGG19. For $\Dct{G}, \Dct{S}\in \Size{c}{h}{w}$, $h\times\!w$ is the spatial resolution and $c$ is the number of channels.  We enforce the consistency between $\Dct{G}$ and $\Dct{S}$ by minimizing their difference measured with the Cross-Entropy (CE) $R_{intra}$ :
\begin{equation}
R_{intra} = -\frac{1}{hw}\sum_{i,j}\sum_{m,n}\mathbf{S}(\Dct{G}[m,n])\log{\mathbf{S}(\Dct{S}[m,n])},
\label{eq:ct}
\end{equation}
where $[m,n]$ is a spatial index of the feature map, and $\mathbf{S}(\cdot)$ denotes the SoftMax function with the input size being  $\mathbb{R}^{c}$. 
{The element-wise CE reflects the negative log-likelihood of distances locally. While other measures are feasible, CE is empirically a good choice in favoring the overall sharpness (see \cref{sec:ablations}).}

\noindent\textbf{Inter-model Distance Distillation} enforces the consistency of changes in low-level characteristics between predictions from different models (orange shading in \cref{fig:overview}). 
Given the two predictions for a single input, \ie $\{\Pred{G}{L}, \Pred{S}{L}\}$ (or $\{\Pred{G}{U}, \Pred{S}{U}\}$), we first calculate their feature distance $\Dap{L}$ (or $\Dap{U}$) to represent the differences in low-level characteristics. However, enforcing consistency on $\Dap{L}$ and $\Dap{U}$ is less straightforward, as the spatial layout of corresponding inputs can differ. Thus, we compute the inter-model distances with the following:
\begin{equation}
\begin{aligned}
    \Dap{L} &= \text{Gram}(\Feat{\Pred{S}{L}} - \Feat{\Pred{G}{L}}), \\
    \Dap{U} &= \text{Gram}(\Feat{\Pred{S}{U}} - \Feat{\Pred{G}{U}}).
\end{aligned}
\end{equation}
$\text{Gram}(\cdot)$ refers to the Gram matrix, which calculates the correlations among vectorized feature maps along the channel dimension. The Gram matrix captures the statistics while collapsing the spatial layout~\cite{gatys2016image}.

For distillation, we ensure the consistency between inter-model distances by minimizing $R_{inter}$, the Frobenius norm between $\Dap{U}$ and $\Dap{L}$:
\begin{equation}
R_{inter} = \sum_{i,j}\Frobenius{\Dap{U} - \Dap{L}},
\label{eq:ap}
\end{equation}
where $\Frobenius{\cdot}$ computes the Frobenius norm. The Frobenius norm of Gram Matrix difference is interpreted in \cite{li2017demystifying,gatys2016image} as measuring distribution discrepancy.

\subsection{Full Method}
\label{Subsec:full_method}
The network $M_S$ is fully optimized with supervised losses for labeled synthetic data $X^L$ and unsupervised losses for real-world data $X^U$.

\noindent\textbf{Supervised Losses} maintain the specialization in synthetic degradation and reduce overfitting to the distillation. With the ground-truth image for prediction $\Pred{S}{L}$, denoted as $\GT$, we implement a combination of supervised loss functions. It includes a wavelet-based loss $L_\text{wv}$, a perceptual loss $L_\text{vgg}$, and generative loss $L_\text{gan}$ for adversarial training:
\begin{equation}
    \mathcal{L}_L(\Pred{S}{L},Y^L) = \alpha_1L_\text{wv}(\Pred{S}{L},\GT) + \alpha_2L_\text{vgg}(\Pred{S}{L},\GT)
    + \alpha_3L_\text{gan}(\Pred{S}{L}),
\label{eq:L_L}
\end{equation}
where $L_\text{vgg}$ and $L_\text{gan}$ are adopted from RealESRGAN~\cite{wang2021real}, with balancing weights $\alpha_1$, $\alpha_2$, and $\alpha_3$. We use but omit writing the discriminator loss for brevity. $L_\text{wv}$ refers to an $\ell_1$-based wavelet loss~\cite{zhong2018joint,deng2019wavelet}:
\begin{equation}
  L_\text{wv}(\Pred{S}{L},\GT) = \sum_{i}\omega_i\|\wvt{\Pred{S}{L}}-\wvt{\GT}\|_1,  
  \label{eq:L_wvt}
\end{equation}
where $\wvt{\cdot}$ extracts the $i$-th wavelet channel, and $\omega_i$ weights its importance. For the wavelet transform, we apply the Haar wavelet~\cite{lepik2014haar} for all output channels.

\noindent\textbf{Unsupervised Losses.} In the absence of ground truth for $\Pred{S}{U}$, we opt to optimize for the consistency outlined in~\cref{eq:ct} and~\cref{eq:ap} as regularizers. To fully utilize the discriminator's knowledge of realness, we also incorporate the generative loss in~\cref{eq:L_L}:
\begin{equation}
\mathcal{L}_U = \lambda_{1}R_{intra}+\lambda_{2}R_{inter} + \lambda_{3}L_{gan}(\Pred{S}{U}),
\label{eq:L_U}
\end{equation}
where $\lambda_1$, $\lambda_2$, and $\lambda_3$ balance the optimization aims. $L_{gan}$ here uses the same discriminator in \cref{eq:L_L}.
The supervised loss in \cref{eq:L_L} and the unsupervised distillation loss in \cref{eq:L_U} are combined as our final optimization objective, expressed as $\mathcal{L}= \mathcal{L}_L + \mathcal{L}_U$. 

\noindent\textbf{Color Correction.} In practice, our approach often leads to color shifts due to the regularization of distances within the feature space. To address this, we rectify the output by normalizing the mean and variance of each color channel with those of the corresponding input channels~\cite{wang2021real}. Further details are provided in Supplementary Sec.\:E. 

\begin{figure}
 \centering
    \includegraphics[width=0.55\linewidth]{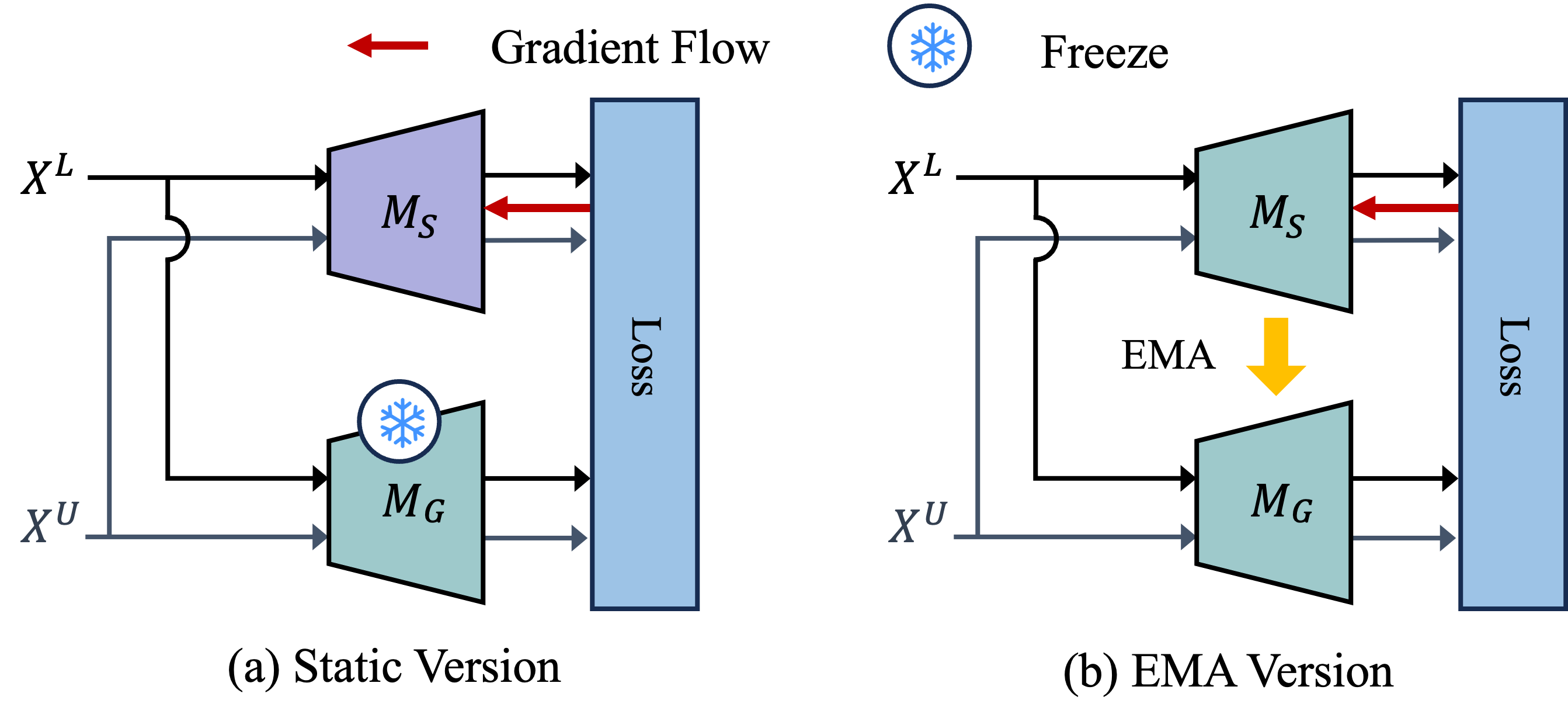}
    \caption{Two initialization options for the generalist and specialist models. (a) The static configuration initializes $M_G$ with a model pre-trained by the complex synthetic pipeline and $M_S$ with a model pre-trained by simple degradation in $X^L$. Weights of $M_G$ are frozen during distillation. (b) Both $M_S$ and $M_G$ are initialized with a pre-trained generalized model. Weights of $M_G$ are the EMA version of $M_S$.}
    \label{fig:SToptions}
\end{figure}

\noindent\textbf{Learning Configurations.} 
Our proposed PDD is a learning framework that can utilize pre-trained networks. It is recommended that $M_G$ is pre-trained with extensive degradations $\mathcal{D}_G$, as in~\cite{wang2021real,zhang2021designing}. 
The \textit{static} configuration (refer to~\cref{fig:SToptions} (a)) initializes $M_S$ with the model pre-trained on specific synthetic degradation ($\mathcal{D}_S$), and $M_G$ is frozen throughout the training process.

Alternative configuration initializes $M_G$ and $M_S$ with the same model pre-trained with $\mathcal{D}_G$ and maintain $M_G$ as an Exponential Moving Average (EMA) version of $M_S$ (see \cref{fig:SToptions} (b)). Despite $M_S$ not being initialized as a specialist, it can readily specialize to the synthetic degradation $\mathcal{D}_S$ through supervised learning in \cref{eq:L_L}. Empirically, the EMA version tends to be more effective, likely because the updated knowledge of real-world data in $M_S$ also benefits $M_G$ through EMA, enabling $M_G$ to yield better references for optimization.

%% file: sec_TS/4_experiments.tex
\section{Experiments}
\label{sec:experiments}
\subsection{Settings}

\noindent\textbf{Datasets.} We experiment on real-world datasets, RealSR~\cite{cai2019toward} and DRealSR~\cite{wei2020component}, and a synthetic dataset without released degradation settings, NTIRE20~\cite{lugmayr2020ntire}. For experiments on RealSR and DRealSR, we synthesize labeled training data $\{X^L\}$ from high-resolution images in the DF2K dataset~\cite{Agustsson_2017_CVPR_Workshops,timofte2017ntire}. For the NTIRE20 dataset, we use the provided target clean images to synthesize low-resolution counterparts. For synthesizing $\{X^L\}$, we use only bicubic interpolation as the default and investigate other options for $\mathcal{D}_S$ in the Supplementary Sec.\:D.b. 
RealSR or DRealSR has multiple data sources, and we optimize our methods for each source separately. For DRealSR, we report results for Panasonic and Olympus sources, as the others do not have sufficient training samples.

\noindent\textbf{Training Details.} We crop patches of size $48\!\times\!48$ as inputs and train the model with the ADAM optimizer~\cite{kingma2014adam} using the settings $\beta_1\!=\!0.9$, $\beta_2\!=\!0.999$, and $\epsilon\!=\!10^{-8}$. The batch size is 16, including equal numbers of labeled and unlabeled inputs. The initial learning rate is $1e-4$ and halved after 25K iterations. The static version needs 50K iterations, and the EMA version is trained for 100K iterations due to the initial specialist underfitting to the synthetic domain. All models are for $\times 4$ super-resolution. Other details are in Supplementary Sec.\:A.

\noindent\textbf{Evaluation.} We evaluate models on testing sets of RealSR, DRealSR, and the validation set of NTIRE20. We crop non-overlapped 512x512 patches from LR images in DRealSR to avoid memory exhaustion. For other datasets, we use the full images. To fully use the provided high-resolution counterparts, we report scores of full-reference metrics, PSNR, SSIM, and LPIPS~\cite{zhang2018unreasonable}. The PSNR and SSIM scores are calculated on the Y-channel of YCbCr format images. Additionally, we follow BSRGAN~\cite{zhang2021designing} and supplement with the no-reference metric NRQM~\cite{ma2017learning} to evaluate the perceptual clearness, as the ground truth in real-world captured datasets~\cite{cai2019toward,wei2020component} also feature some blur. We use the implementation in IQA-PyTorch\footnote{https://github.com/chaofengc/IQA-PyTorch} for all metrics.
It is important to consider both fidelity and perceptual scores; fidelity metrics are worse at detecting blurriness while perceptual metrics are not sensitive to high-frequency artifacts~\cite{zhang2022perception,ma2020structure,maggioni2023tunable,park2023perception}.

\begin{table}[t]
\caption{$M_G$ and $M_S$(static) columns show \textit{initial} performances on the bicubic-interpolated data $\{X^L\}$ from Set14~\cite{zeyde2012single} and $\{X^U\}$ with unknown degradation from NTIRE20~\cite{lugmayr2020ntire}. $M_S$(EMA) column shows the performance of $M_S$ in the EMA version on $\{X^L\}$ \textit{after} training process.}
    \centering
    \resizebox{0.85\linewidth}{!}{
    \begin{tabular}{c|c|ccc|c|c|cc}
    Domain&Metric & $M_G$ & $M_S$(static) & $M_S$(EMA)&Domain&Metric & $M_G$ & $M_S$(static)\\
    \hline
     \multirow{2}{*}{$\{X^L\}$}&PSNR$\uparrow$& 23.98 &24.81 &26.26 &\multirow{2}{*}{$\{X^U\}$} &PSNR$\uparrow$& 25.08 & 19.82\\
         &LPIPS$\downarrow$&0.2349 &0.1337 &0.1471&&LPIPS$\downarrow$&0.2504 &0.7552
    \end{tabular}}
    \label{tab:assumption}
\end{table}
\input{Figures/Experiments/pddm}

\subsection{Effectiveness of PDD}
\label{subsec:PDDM}

\noindent\textbf{Comparisons.} We compare the two versions of our method in \cref{fig:SToptions} with the naive distillation defined in \cref{eq:baseline} regarding their improvements over the generalist network $M_G$. We choose RealESRGAN~\cite{wang2021real} as the pre-trained generalist model and ESRGAN\cite{wang2018esrgan}, a standard SISR model, as the pre-trained specialist for the naive distillation and the static version in \cref{fig:SToptions} (a). \cref{tab:assumption} shows the initial performances of the generalist and specialist for $\{X^L\}$ and $\{X^U\}$. As we assumed in \cref{Subsec:preliminary}, the specialist performs better than the generalist for $\{X^L\}$ while $M_G$ generalizes better for $\{X^U\}$.
For the EMA version in \cref{fig:SToptions} (b), $M_S$ as well as $M_G$ is initialized with RealESRGAN. Through the supervised learning in \cref{eq:L_L}, $M_S\text{(EMA)}$ shows specialized performance for labeled data $\{X^L\}$.

\cref{fig:cmp2T} plots the improvements over the pretrained generalist for Naive Distillation (ND), Static, and EMA versions of our method. Compared to the generalist, ND dramatically drops in perceptual scores (LPIPS, NRQM) despite maintaining or improving the PSNR.
Both versions of our method perform better than ND; compared to the generalist, our static version improves PSNR significantly and at least one perceptual score for each data domain. The EMA version improves all metrics with the highest magnitudes. 

\noindent\textbf{Change in Low-Level Characteristics.} 
We extract low-level features for visualization following \cite{liu2021discovering}. \cref{subfig:initial_MS} visualize predictions generated by the bicubic specialist (ESRGAN), which has separate distributions. \cref{subfig:ultimate_MS} verifies the predictions from the specialist after applying our method (static version), where the separated distributions are pushed closer. 
Visualizing low-level differences between images with arbitrary content is challenging~\cite{liu2021discovering}. As such, we also quantitatively compare the KL divergence between the low-level features. As shown in \cref{fig:delta_KL}, applying our method decreases the KL divergence between labeled (DIV2K-bicubic) and unlabeled predictions.
Details of the visualization and KL-divergence calculations are in Supplementary Sec. B.

\input{Figures/Experiments/dist_shift}

\subsection{Comparisons with Other Methods}
\label{subsec:sota_cmp}
\begin{figure}[t]
    \begin{minipage}{0.45\textwidth}
    \centering
    \includegraphics[width=0.7\linewidth]{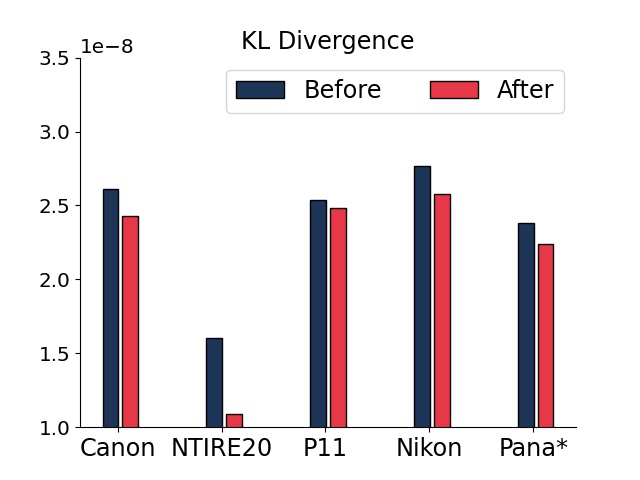}
    \captionof{figure}{KL divergence between specialist's unlabeled and labeled predictions. The KL divergence between labeled and unlabeled predictions is lower after applying our method (\textcolor[HTML]{e73847}{red}) than that in the specialist before adaption (\textcolor[HTML]{1d3557}{blue}).}
    \label{fig:delta_KL}
    \end{minipage}
    \hfill
    \begin{minipage}{.5\textwidth}
    \centering
    \includegraphics[width=\linewidth]{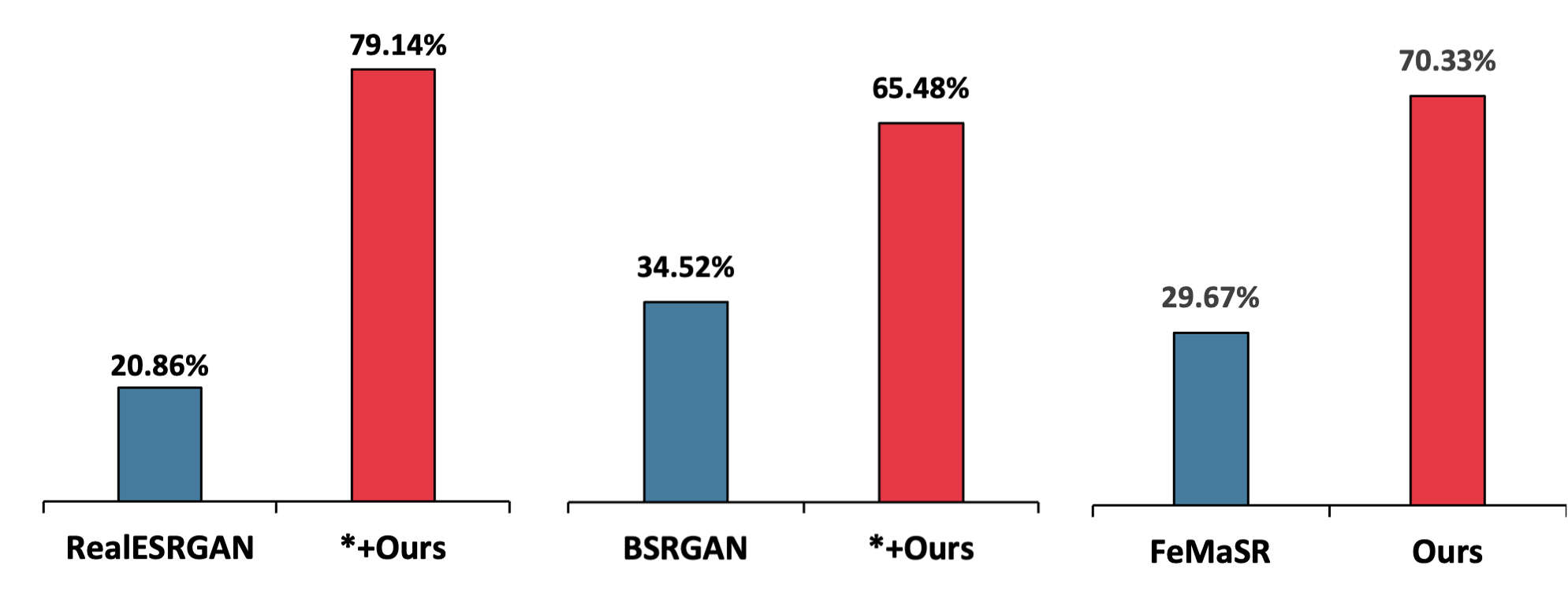}
    \captionof{figure}{User study on outputs of 30 real-world images. We recruit 30 evaluators for each image. Each vote is for a better quality within a pair of options. We compared our method to RealESRGAN and BSRGAN by using them as pre-trained models; FeMaSR is compared to RealESRGAN+ours. Our method gains at least 65\% votes.}
    \label{fig:user_study}
    \end{minipage}
\end{figure}

\input{tables/results}
\textbf{Quantitative Comparison.} We consider two variants of our method by applying the EMA version on RealESRGAN~\cite{wang2021real} and BSRGAN~\cite{zhang2021designing}. In \cref{tab:results}, we compare their performances with other state-of-the-art RWSR methods, including FeMaSR~\cite{chen2022real}, EDAN~\cite{liang2022efficient}, DASR~\cite{wei2021unsupervised}, DAN~\cite{huang2020unfolding}, RealESRGAN~\cite{wang2021real}, BSRGAN~\cite{zhang2021designing}, and TG~\cite{zhang2024real}, which released code and model weights that enable reproduction.
We regenerate results to evaluate these methods, while Supplementary Sec. C discusses related but not directly comparable methods.

As shown in \cref{tab:results}, EDAN, DAN, and DASR perform strongly on the fidelity metrics (PSNR, SSIM) but are poor perceptually, especially on the NRQM score. The qualitative comparison highlights the blurry effects. Our method enhances RealESRGAN and BSRGAN across all metrics and demonstrates superior performance compared to TG when both are applied to RealESRGAN. For FeMaSR, we achieve competitive perceptual scores and significantly surpass their fidelity scores.
\input{Figures/Experiments/qualitative_cmp}

\noindent\textbf{User Study} is conducted by presenting two reconstructions of the same LR image and asking for the better one. We compare the EMA version of our method with methods RealESRGAN, BSRGAN, and FeMaSR.
Each comparison is conducted on 30 different real-world results among 30 evaluators.
As shown in \cref{fig:user_study}, at least 65\% evaluators vote for our method in each comparison. Although quantitatively competing, FeMaSR yields unpleasant artifacts to human eyes, further shown in qualitative comparison.

\noindent\textbf{Qualitative Comparison} is in \cref{fig:qualitative_cmp}. FeMaSR~\cite{chen2022real} produces high-frequency artifacts, while outputs from DASR~\cite{wei2021unsupervised} suffer blur or artifacts. Compared to RealESRGAN, our method predicts sharper patterns without introducing distortion. More comparisons are in Supplementary Sec.\:F.

\subsection{Ablations}
\label{sec:ablations}

\begin{figure}[t]
  \centering
  \includegraphics[width=0.7\linewidth]{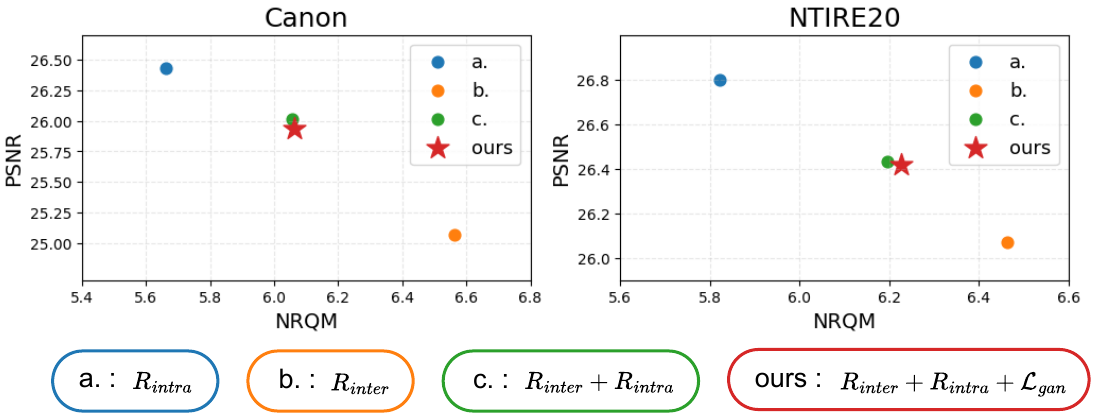}
  \caption{Ablations of loss terms for distillation in \cref{eq:L_U}. Using one of $R_{intra}$ (\textcolor{blue}{\textit{blue}}) and $R_{inter}$ (\textcolor{orange}{\textit{orange}}) bias fidelity or perceptual quality (a.-b.), while the combination of them (c.-d.) achieves a balanced improvement (see \textcolor{teal}{\textit{green}} and \textcolor{red}{\textit{red}} dots).}
  \label{fig:abl_loss}
\end{figure}

\noindent\textbf{Ablation on loss terms} reveals the effectiveness of each term in \cref{eq:L_U}.
All experiments take RealESRGAN as the pre-trained model and follow the EMA implementation of our method in \cref{fig:SToptions} (b). \cref{fig:abl_loss} shows that $R_{intra}$ and $R_{inter}$ relate differently to fidelity and perception. Their combination achieves balanced improvements in both directions, and the $L_\text{gan}$ increases pattern sharpness with a slight trade-off in PSNR.
Controlling the ratio between weights of $R_{intra}$ and $R_{inter}$ results in fidelity and perceptual quality trade-offs. We choose their balance in \cref{subsec:sota_cmp} and refer to Supplementary Sec.\:D.a for more discussion.

\input{tables/abla_collection}

\noindent\textbf{Effectiveness of EMA} is further validated by comparing to a \textit{SingleFixed} version, which also uses RealESRGAN as the pre-trained model for both $M_S$ and $M_G$ but fixes the $M_G$ during training. As shown in \cref{tab:abla_collection}, updating the weights of the generalist network through EMA enables higher performance in both fidelity and perception.

\noindent\textbf{Choice of $R_{intra}$ measurement.} \cref{eq:ct} use Cross-Entropy (CE) for measuring the difference between $\Dct{S}$ and $\Dct{G}$. Here, we alternate the CE to $\ell_1$ distance without changing the ratio between weights of $R_{intra}$ and $R_{inter}$. As shown in \cref{tab:abla_collection}, CE favors the overall sharpness while $\ell_1$ focuses more on fidelity.
Although $\ell_1$ is a feasible choice, we use CE in our method due to the importance of sharpness for human eyes~\cite{ledig2017photo}.

%% file: Figures/Experiments/pddm.tex
\begin{figure}[tb]
\centering
    \begin{subfigure}[t]{0.325\linewidth}
    \includegraphics[width=\textwidth]{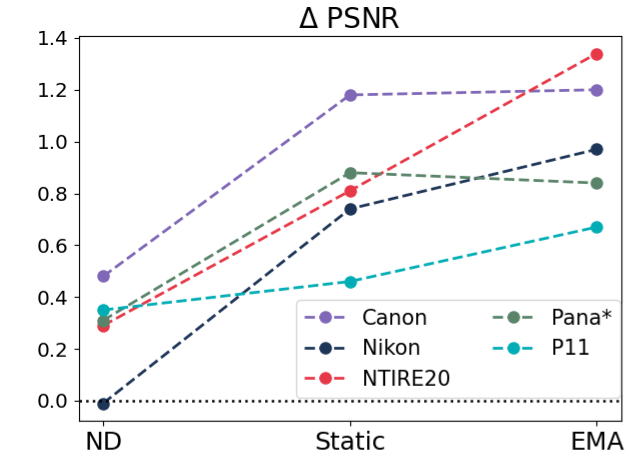}
    \end{subfigure}
    \begin{subfigure}[t]{0.325\linewidth}
    \includegraphics[width=\textwidth]{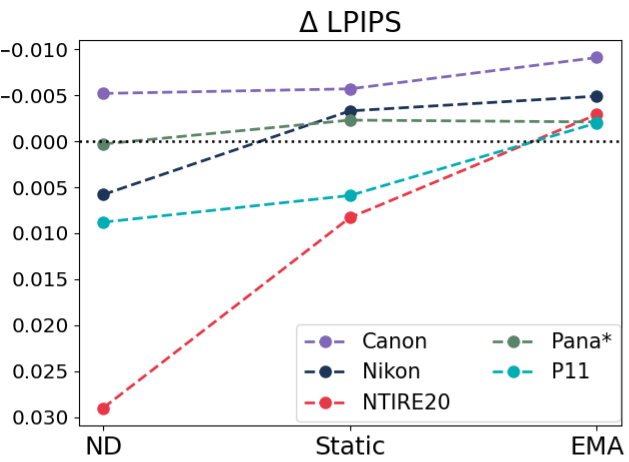}
    \end{subfigure}
    \begin{subfigure}[t]{0.325\linewidth}
    \includegraphics[width=\textwidth]{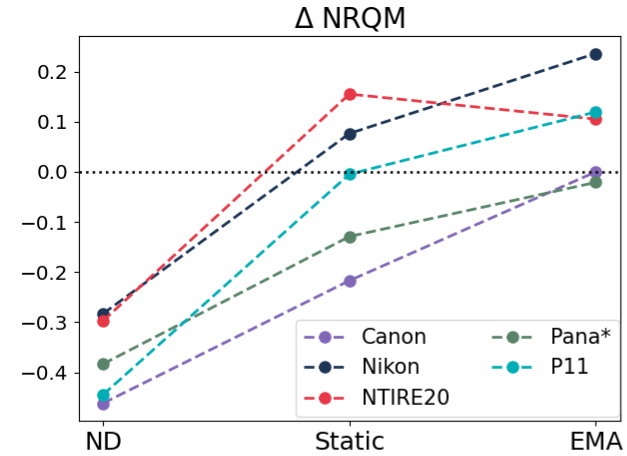}
    \end{subfigure}
\caption{{Improvements over the Generalist on five unlabeled data domains, where a lower LPIPS score is better. ND improves fidelity scores (PSNR) but dramatically drops perceptual scores ({LPIPS and NRQM}). Both versions of our method achieve better improvements than ND for all reported metrics. For each domain, the Static version improves at least one of the perceptual metrics; the EMA version improves all.}}
\label{fig:cmp2T}
\end{figure}

%% file: Figures/Experiments/dist_shift.tex
\begin{figure}[t]
    \centering
    \InsertSubfigWithCapWithLabel{0.27}{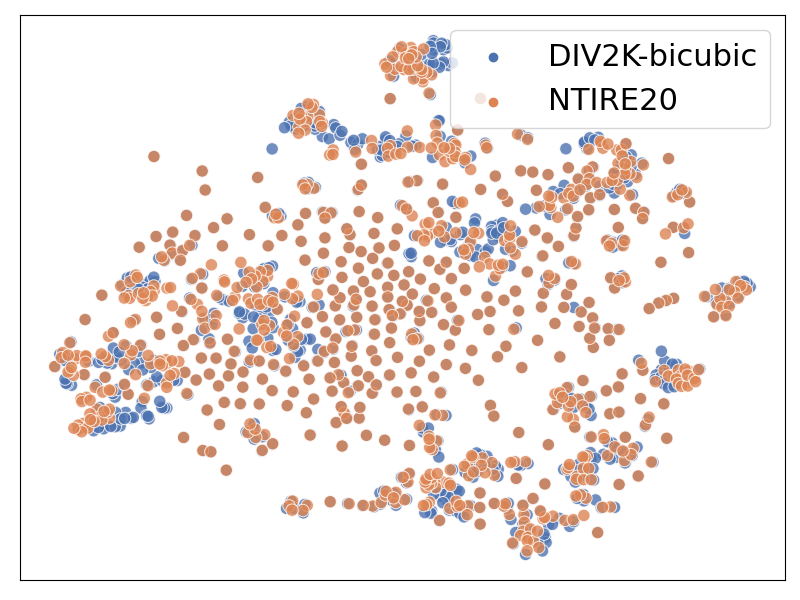}{Generalist}{subfig:dist_MG}
    \InsertSubfigWithCapWithLabel{0.27}{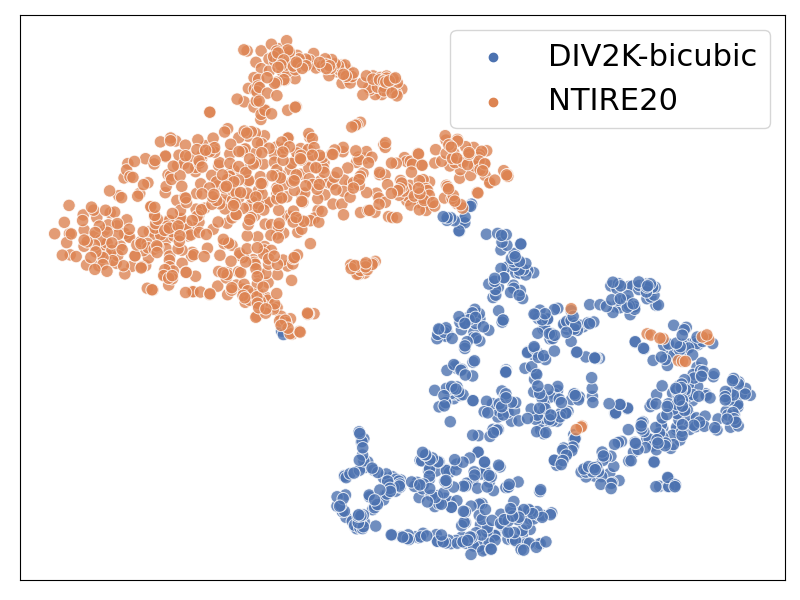}{Bicubic Specialist}{subfig:initial_MS}
    \InsertSubfigWithCapWithLabel{0.27}{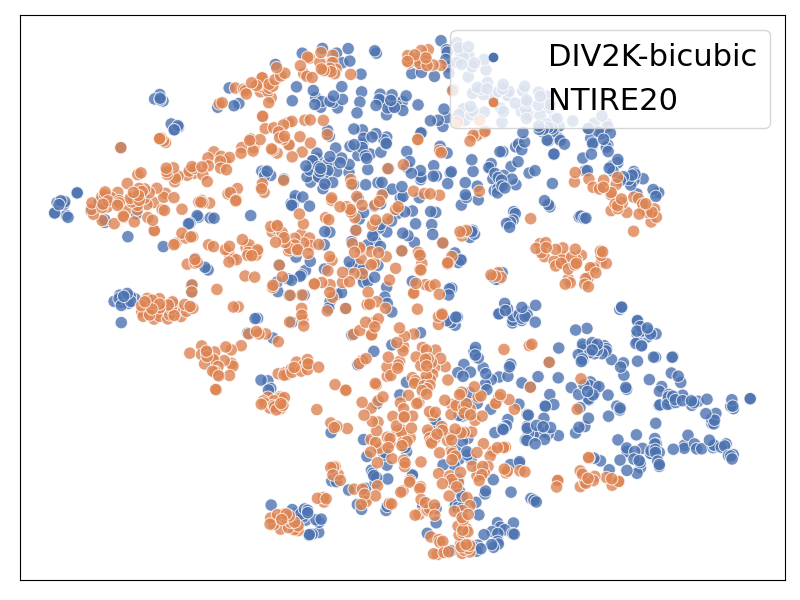}{Adapted Specialist}{subfig:ultimate_MS}
    \caption{Visualization of low-level features for predictions of DIV2K~\cite{Agustsson_2017_CVPR_Workshops} (bicubic) and NTIRE20 (unknown) following~\cite{liu2021discovering} (a) Generalist's predictions for the two domains has overlapped distribution. (b) The predictions from the specialist that only adept in bicubic interpolation have separated distributions. (c) After applying our method (static version), predictions for two domains are pushed close.}
    \label{fig:dist_shift}
\end{figure}

%% file: tables/results.tex
\begin{table*}[!h]
    \centering
    \caption{{Quantitative comparison} with state-of-the-art methods. DAN, EDAN, and DASR are separately listed as they favor fidelity metrics but yield poor perceptual scores. For other methods, \textbf{bold} and \underline{underline} mark the best and second-best scores. Subscripts in the last two columns show their difference from the corresponding pre-trained generalist, where improvement is colored \textcolor{red}{red} and drop is \textcolor{blue}{blue}. }
    \resizebox{\linewidth}{!}{
    \begin{tabular}{c|c|c c c| c c  c c c c}
    \hline
    \hline
        \multirow{2}{*}{Dataset} & \multirow{2}{*}{Metric} & DAN & EDAN & DASR & FeMaSR & RealESRGAN & BSRGAN&RealESRGAN &RealESRGAN & BSRGAN\\
         & &\cite{huang2020unfolding}&\cite{liang2022efficient} &\cite{wei2021unsupervised} &\cite{chen2022real} &\cite{zhang2021designing} & \cite{wang2021real} & +TG\cite{zhang2024real} & +Ours & +Ours\\
         \hline
         \multirow{5}{*}{\makecell{NTIRE20\\-Valid}}& PSNR$\uparrow$ &26.83 & 26.59 & - &23.54 & 25.08 &25.44&25.56 &\improve{\Best{26.42}}{+1.340} &\improve{\Second{26.37}}{+0.930} \\
         & SSIM$\uparrow$ &0.7171 &0.7400  & - &0.6665 &0.7061  &0.6984 &0.7193&\improve{\Best{0.7297}}{+0.024} & \improve{\Second{0.7238}}{+0.025} \\
         & LPIPS$\downarrow$ &0.5747 & 0.2678 & - &\Best{0.2360} &0.2504  &0.2645&\Second{0.2420} &\improve{0.2475}{-0.003} & \improve{0.2595}{-0.005}  \\
         & NRQM$\uparrow$ &4.8784 & 5.8481 & - &\Best{6.6245} & 6.1213 &6.2779&5.7926 &\improve{6.2263}{+0.105} & \improve{\Second{6.2781}}{+0.0002}  \\
         \hline
         \multirow{5}{*}{\makecell{RealSR\\-Canon}}& PSNR$\uparrow$& 26.67&26.36  & 26.71 &24.29 & 24.74 &25.57& 25.03&\improve{\Second{25.94}}{+1.200} & \improve{\Best{26.13}}{+0.560} \\
         & SSIM$\uparrow$ &0.7740 &0.7774  & 0.7782 &0.7460 & 0.7634 &0.7683&\Second{0.7720} &\improve{\Best{0.7733}}{+0.010} & \drop{0.7626}{-0.006} \\
         & LPIPS$\downarrow$ &0.4095 & 0.3026 & 0.2507 &0.2809 & 0.2607 &0.2573&0.2615 & \improve{\Best{0.2516}}{-0.009}& \improve{\Second{0.2517}}{-0.006} \\
         & NRQM$\uparrow$ &3.0407 & 3.6360 &3.9383  &6.0151 & \Second{6.0649} &6.0293&5.4250 &\drop{\Second{6.0647}}{-0.0002} & \improve{\Best{6.1323}}{+0.103} \\
         \hline
         \multirow{5}{*}{\makecell{RealSR\\-Nikon}}& PSNR$\uparrow$ &26.29 & 25.37 &25.63  &23.94 & 24.31 & 24.79& 24.56&\improve{\Best{25.28}}{+0.970} & \improve{\Second{25.07}}{+0.280} \\
         & SSIM$\uparrow$ &0.7533 & 0.7460 & 0.7473 &0.7097 &0.7406&0.7334 &\Second{0.7406} &\improve{\Best{0.7470}}{+0.014} & \drop{0.7294}{-0.001} \\
         & LPIPS$\downarrow$ &0.4133 & 0.3200 & 0.2785 &0.3046 & 0.2851 &\Second{0.2797}&0.2949 &\improve{0.2802}{-0.005} & \improve{\Best{0.2713}}{-0.008} \\
         & NRQM$\uparrow$ &3.2056 & 4.4872 & 4.4952 &\Second{5.9488} & 5.6685 &5.9387&5.1942 &\improve{5.9044}{+0.236} & \improve{\Best{6.1276}}{+0.189} \\
         \hline
         \multirow{5}{*}{\makecell{DRealSR\\-Panasonic}}& PSNR$\uparrow$ & 28.99&28.12  & - & 25.02&27.03  &26.93&27.51 & \improve{\Best{27.87}}{+0.840}& \improve{\Second{27.69}}{+0.760} \\
         & SSIM$\uparrow$ &0.8123 &0.8006  & - &0.7118 &0.7782  &0.7621&\Second{0.7843} &\improve{\Best{0.7939}}{+0.016} & \improve{0.7752}{+0.013} \\
         & LPIPS$\downarrow$ &0.4235 &0.2747  &- &0.3088 &0.2703  &0.2820&0.2836 & \improve{\Best{0.2682}}{-0.002}&\improve{\Second{0.2687}}{-0.013}  \\
         & NRQM$\uparrow$ &2.8131 & 4.5915 & - &\Best{6.0342} & 5.3751 &\Second{5.5865} &4.6483&\drop{5.3540}{-0.021} & \drop{\Second{5.5608}}{-0.026} \\
         \hline
         \multirow{5}{*}{\makecell{DRealSR\\-Olympus}}& PSNR$\uparrow$ &28.74 &27.85  &-  &24.38 & 26.73 &26.49 &\Best{27.43}& \improve{\Second{27.40}}{+0.670}& \improve{27.25}{+0.764} \\
         & SSIM$\uparrow$&0.8063 &0.7979  &-  &0.6691 & 0.7667 &0.7568&\Second{0.7843} &\improve{\Best{0.7846}}{+0.018} & \improve{0.7635}{+0.007} \\
         & LPIPS$\downarrow$ &0.4595 & 0.3555 &-  &0.4007 & \Second{0.3159} &0.3283 &0.3416&\improve{\Best{0.3139}}{-0.002} & \improve{0.3268}{-0.001} \\
         & NRQM$\uparrow$ &2.7961 & 3.7989 &-  &\Best{6.2590} & 5.2530 &5.5376&4.1459 &\improve{5.3722}{+0.120} & \improve{\Second{5.5957}}{+0.058} \\
         \hline
    \end{tabular}}
    \label{tab:results}
\end{table*}

%% file: Figures/Experiments/qualitative_cmp.tex
\begin{figure}
    \centering
    \InsertSubfig{0.15}{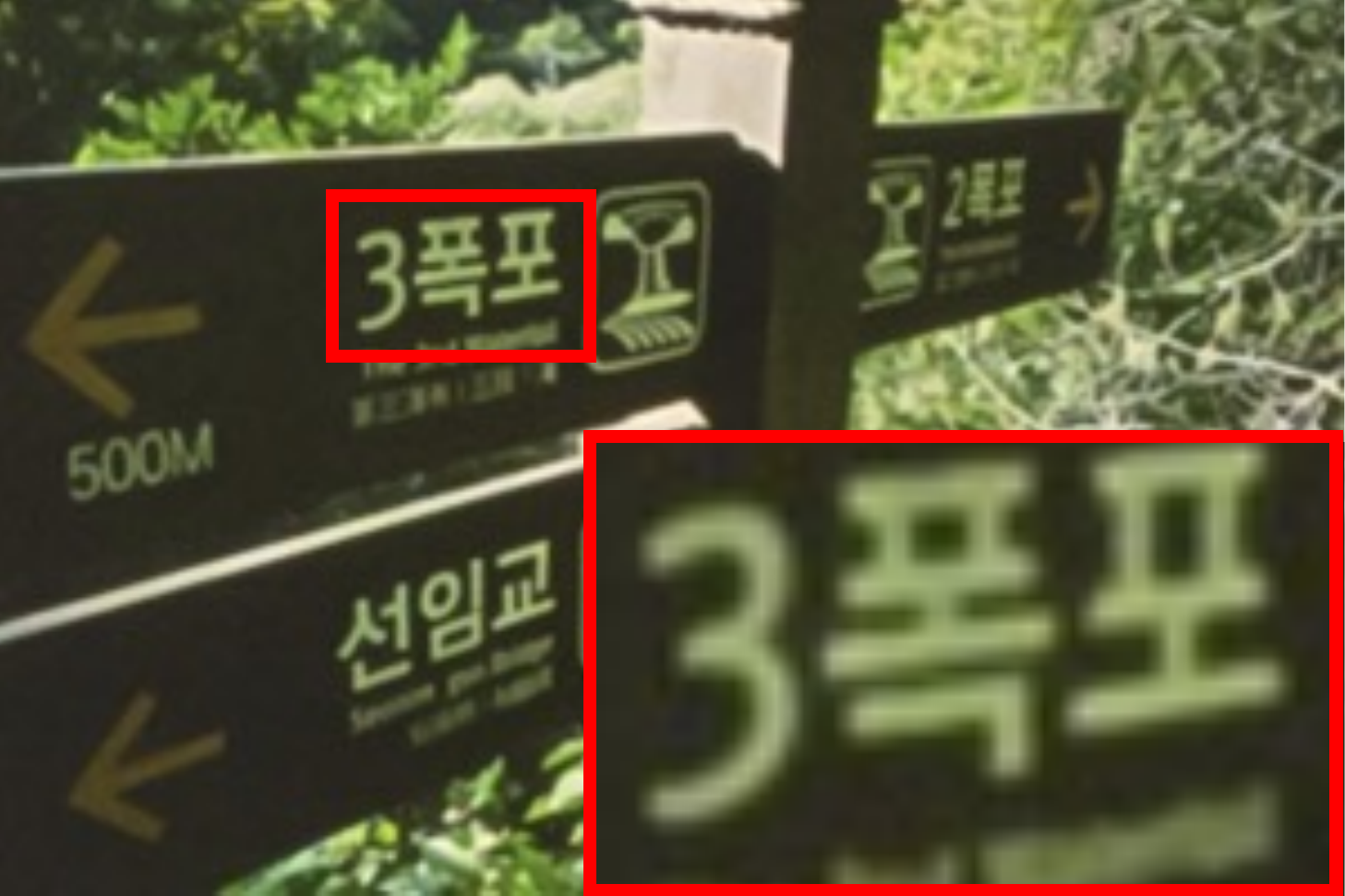}
    \InsertSubfig{0.15}{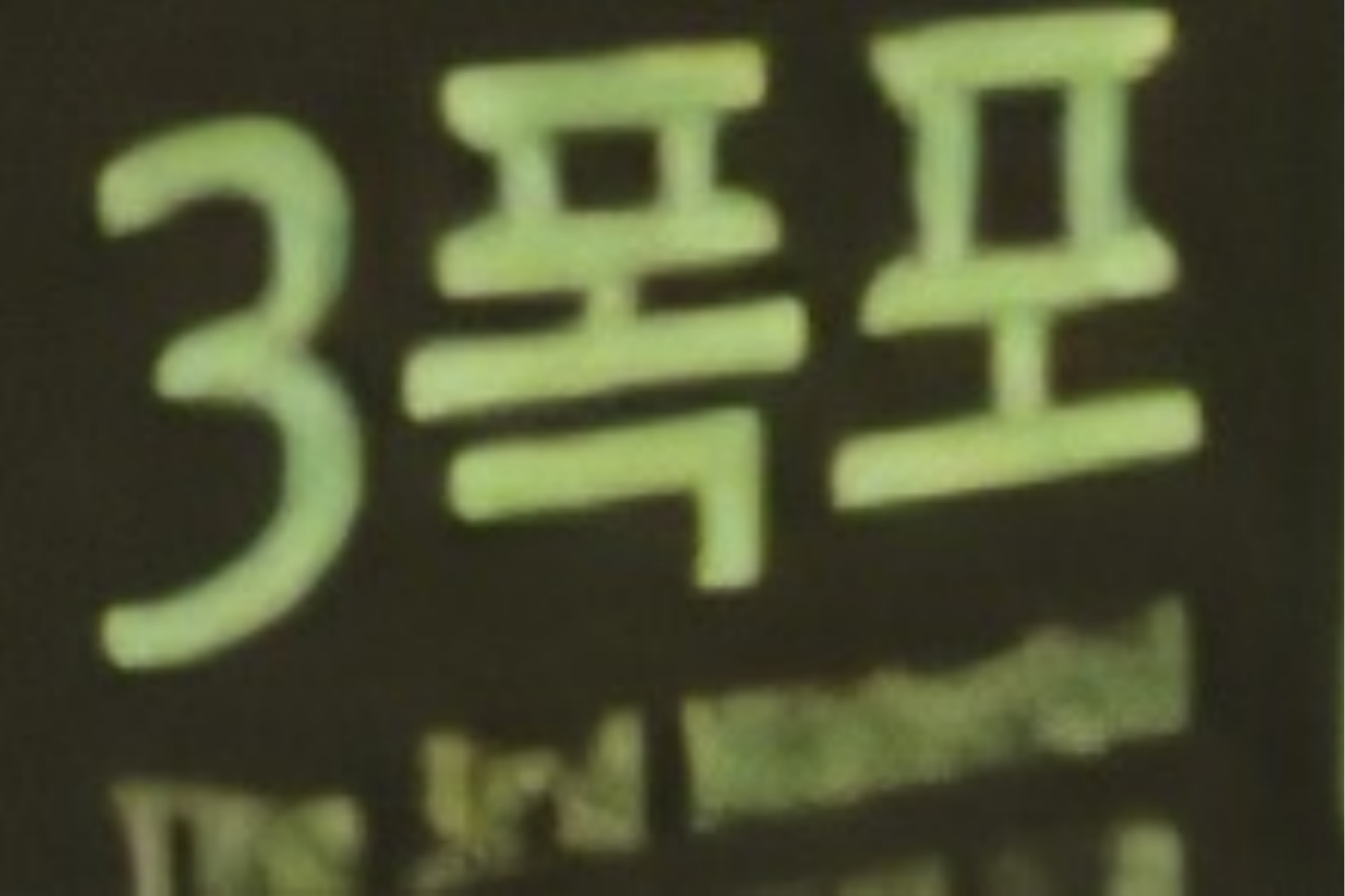}
    \InsertSubfig{0.15}{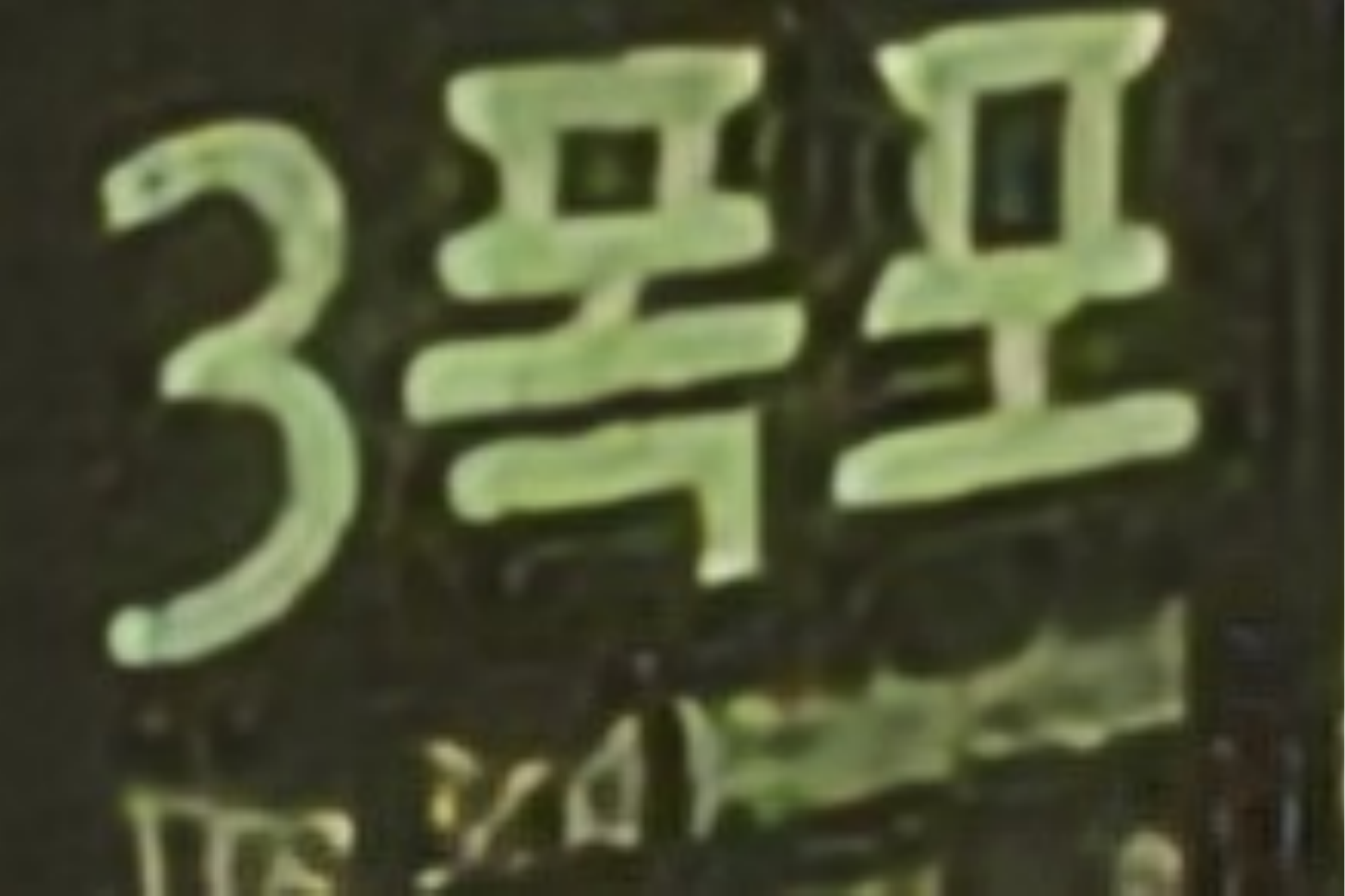}
    \InsertSubfig{0.15}{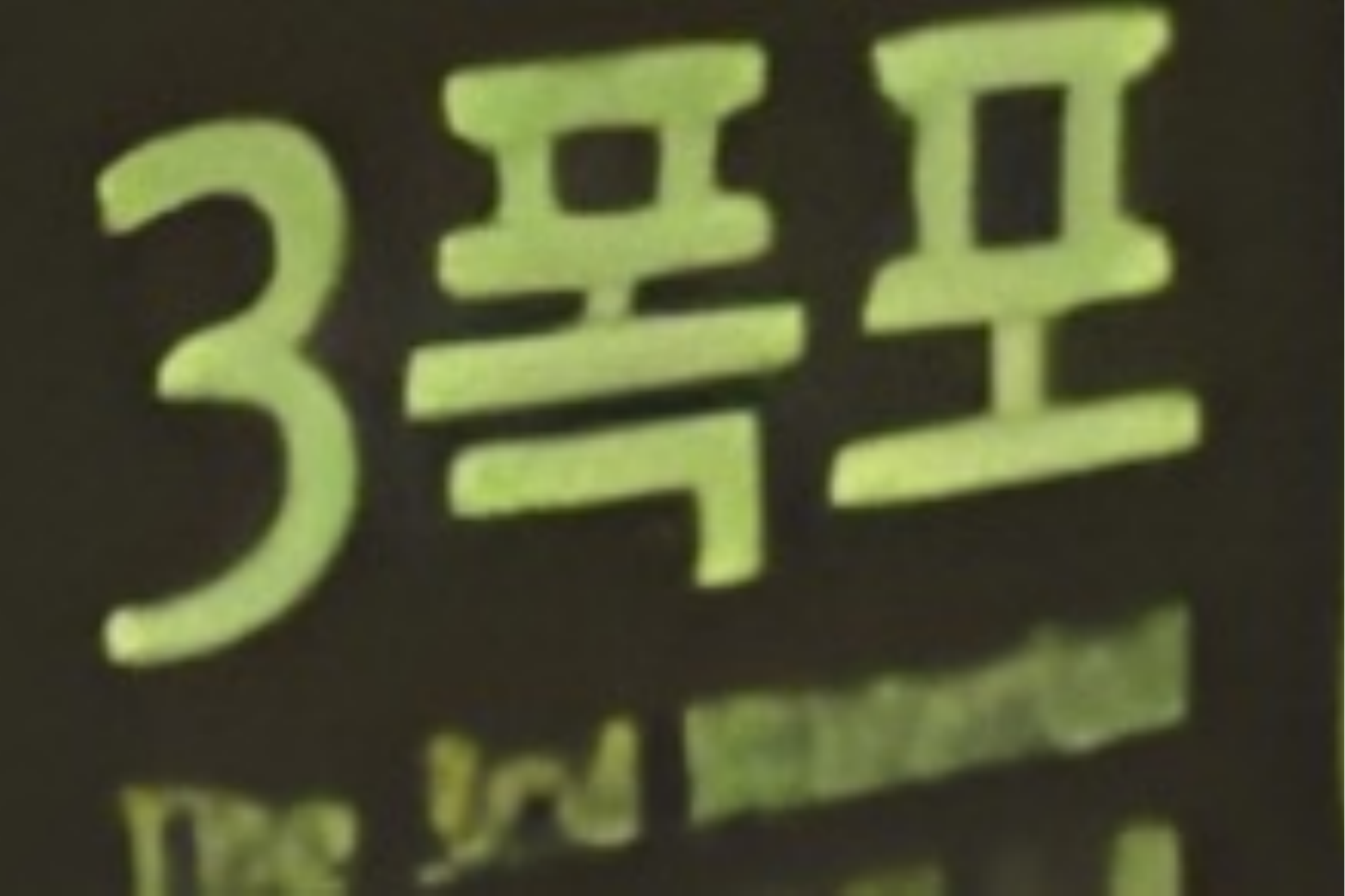}
    \InsertSubfig{0.15}{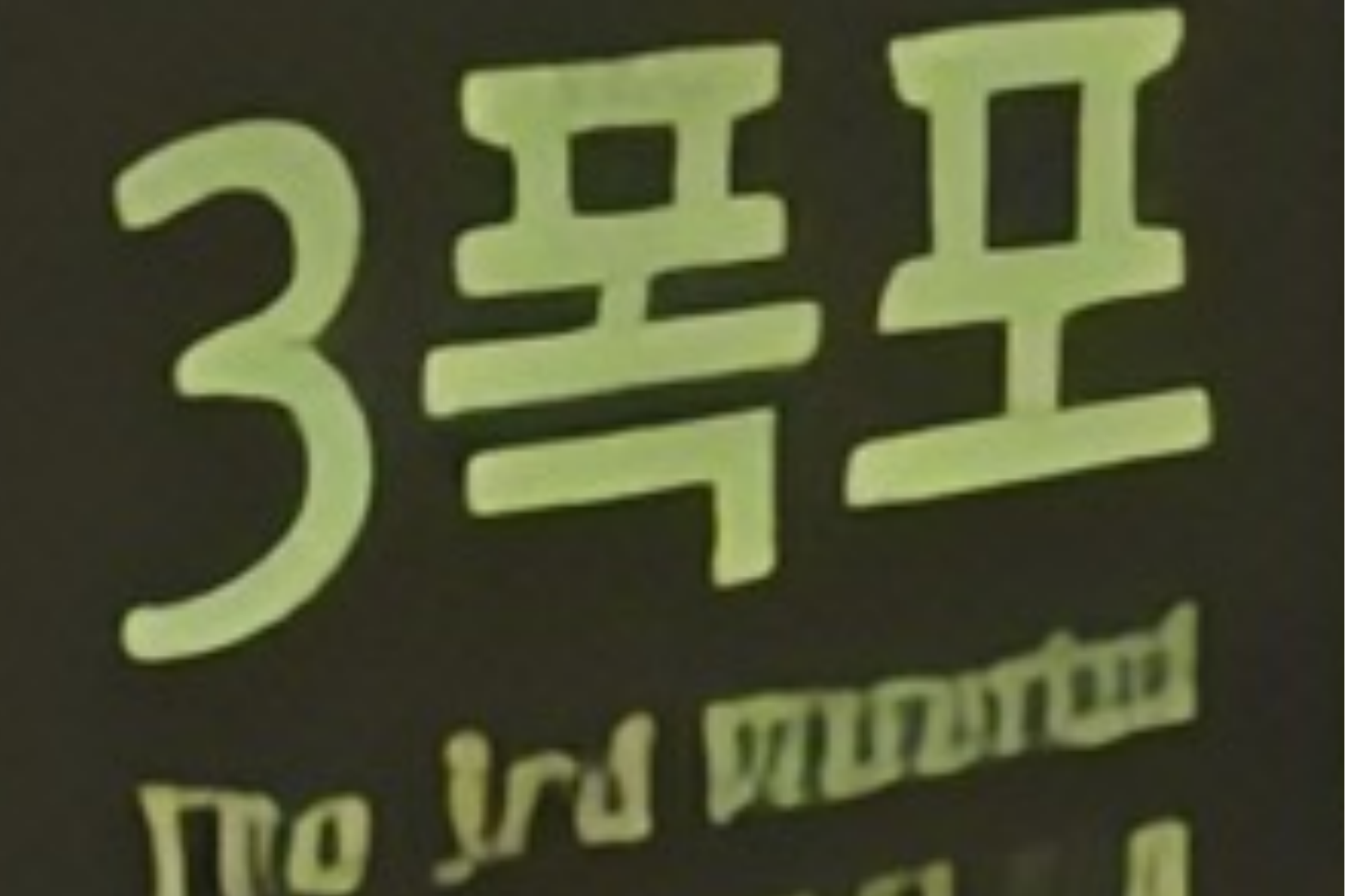}
    \InsertSubfig{0.15}{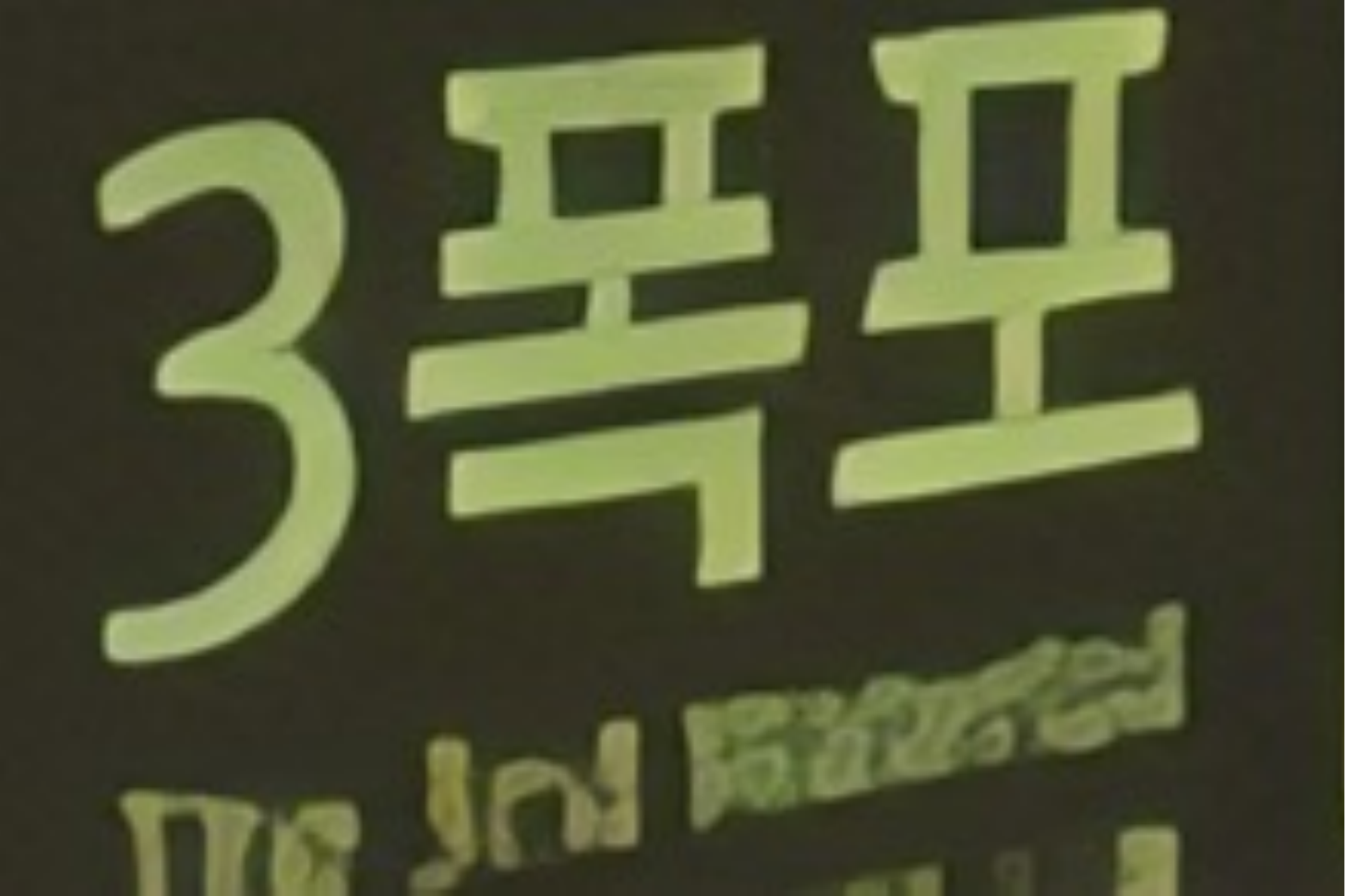}

    \InsertSubfig{0.15}{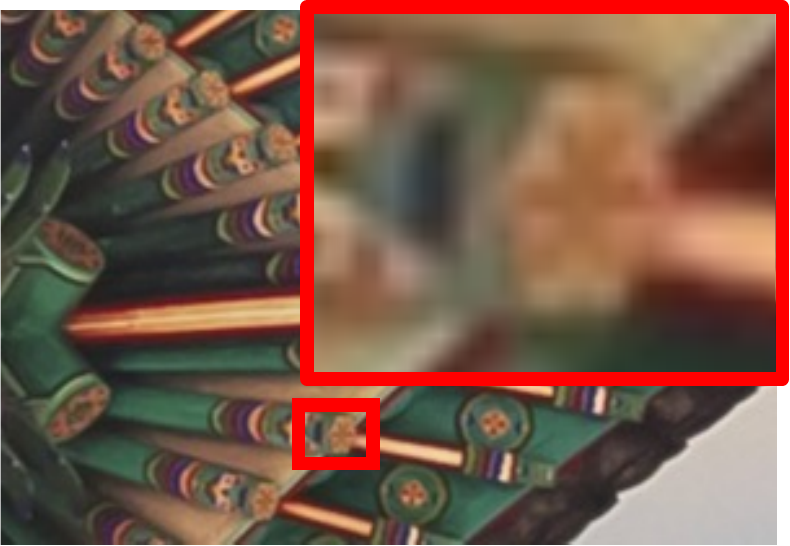}
    \InsertSubfig{0.15}{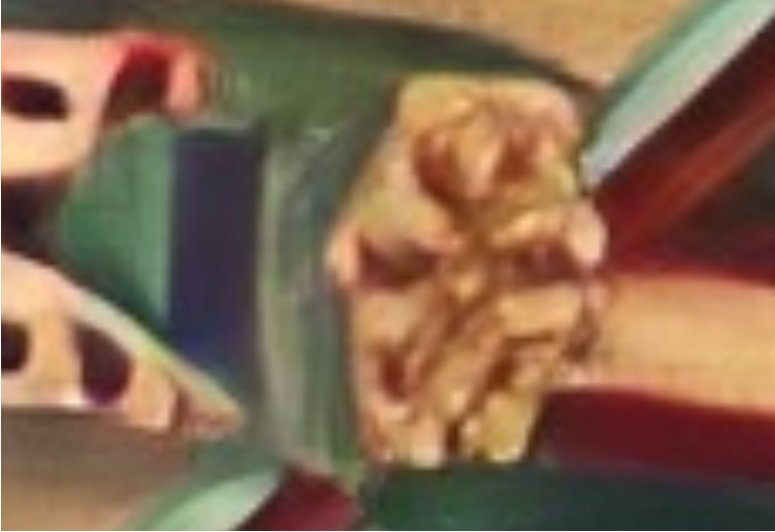}
    \InsertSubfig{0.15}{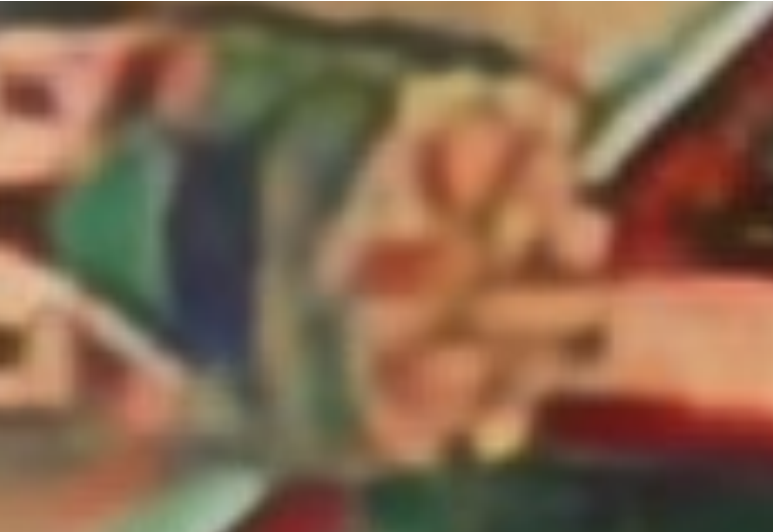}
    \InsertSubfig{0.15}{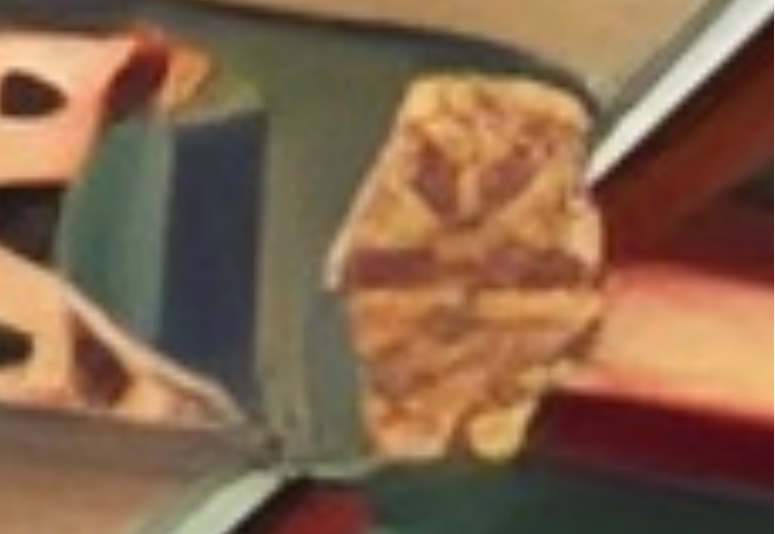}
    \InsertSubfig{0.15}{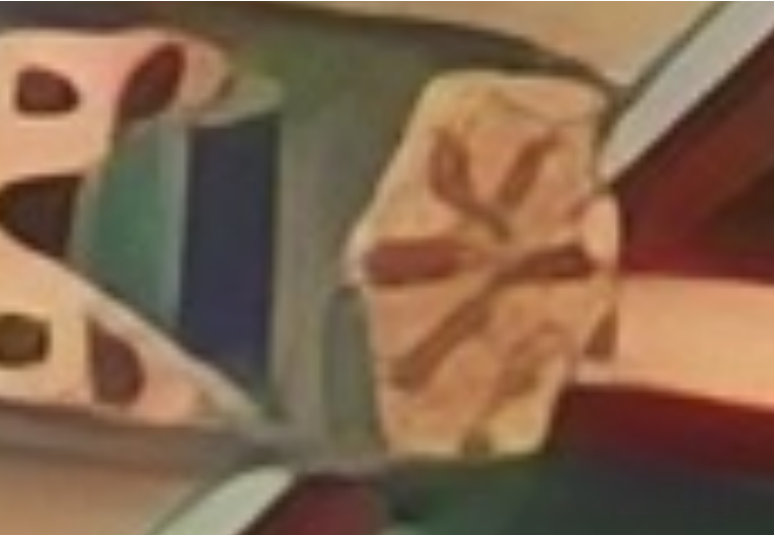}
    \InsertSubfig{0.15}{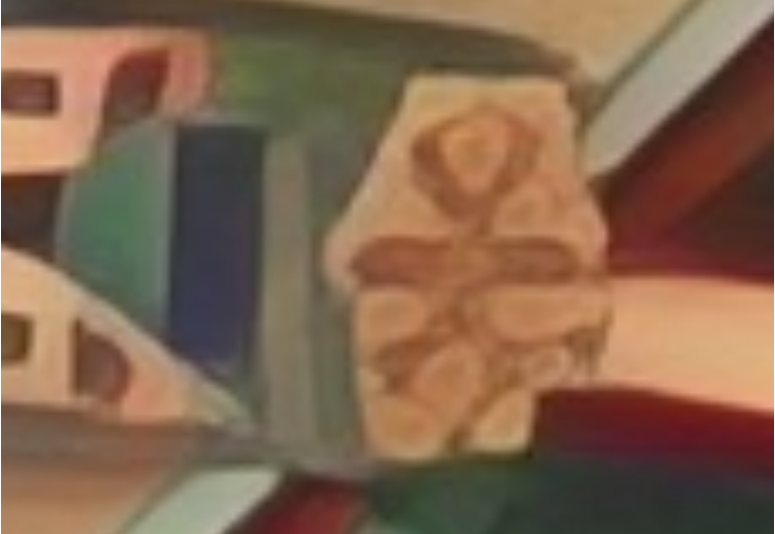}
    
    \InsertSubfig{0.15}
    {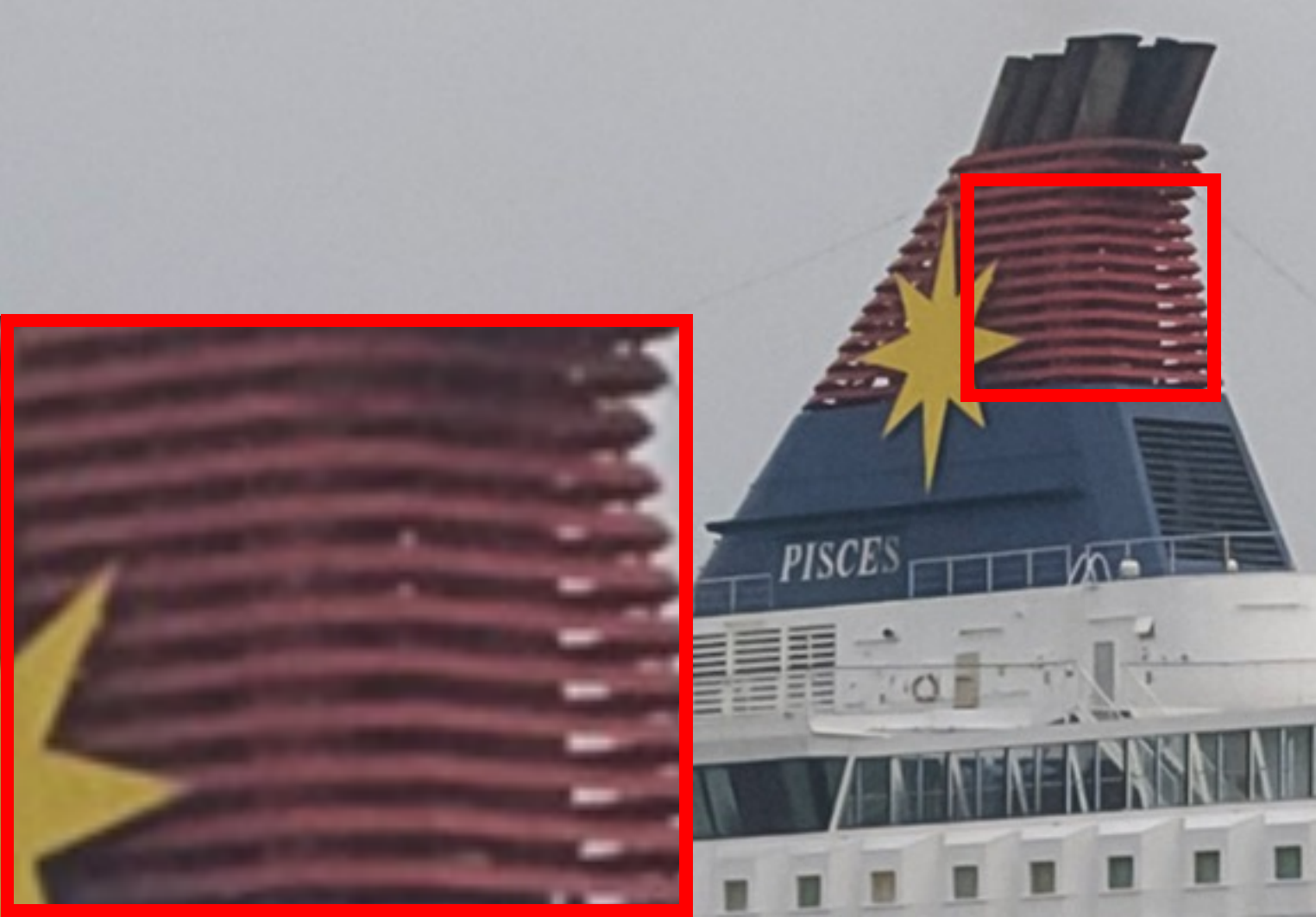}
    \InsertSubfig{0.15}{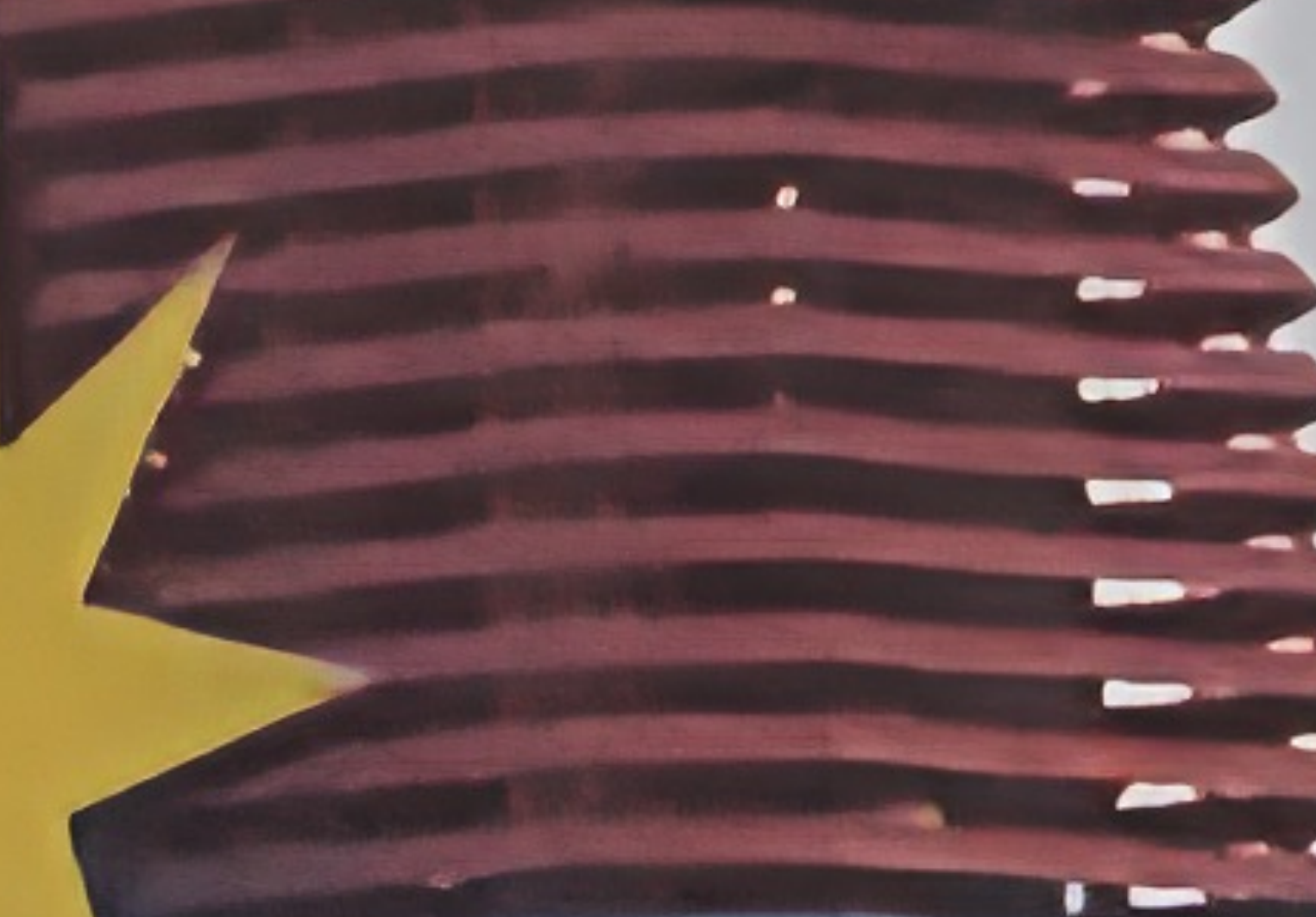}
    \InsertSubfig{0.15}{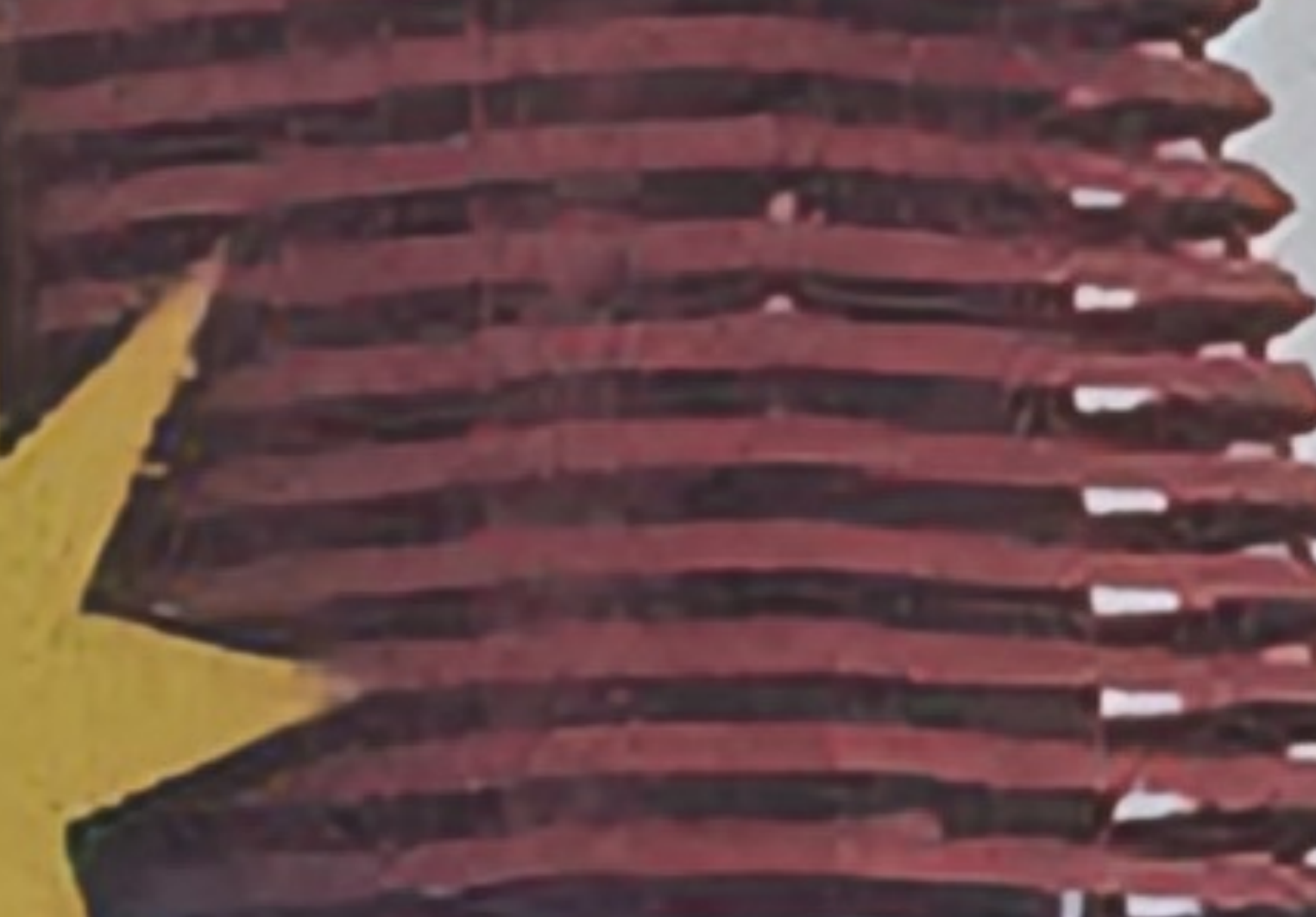}
    \InsertSubfig{0.15}{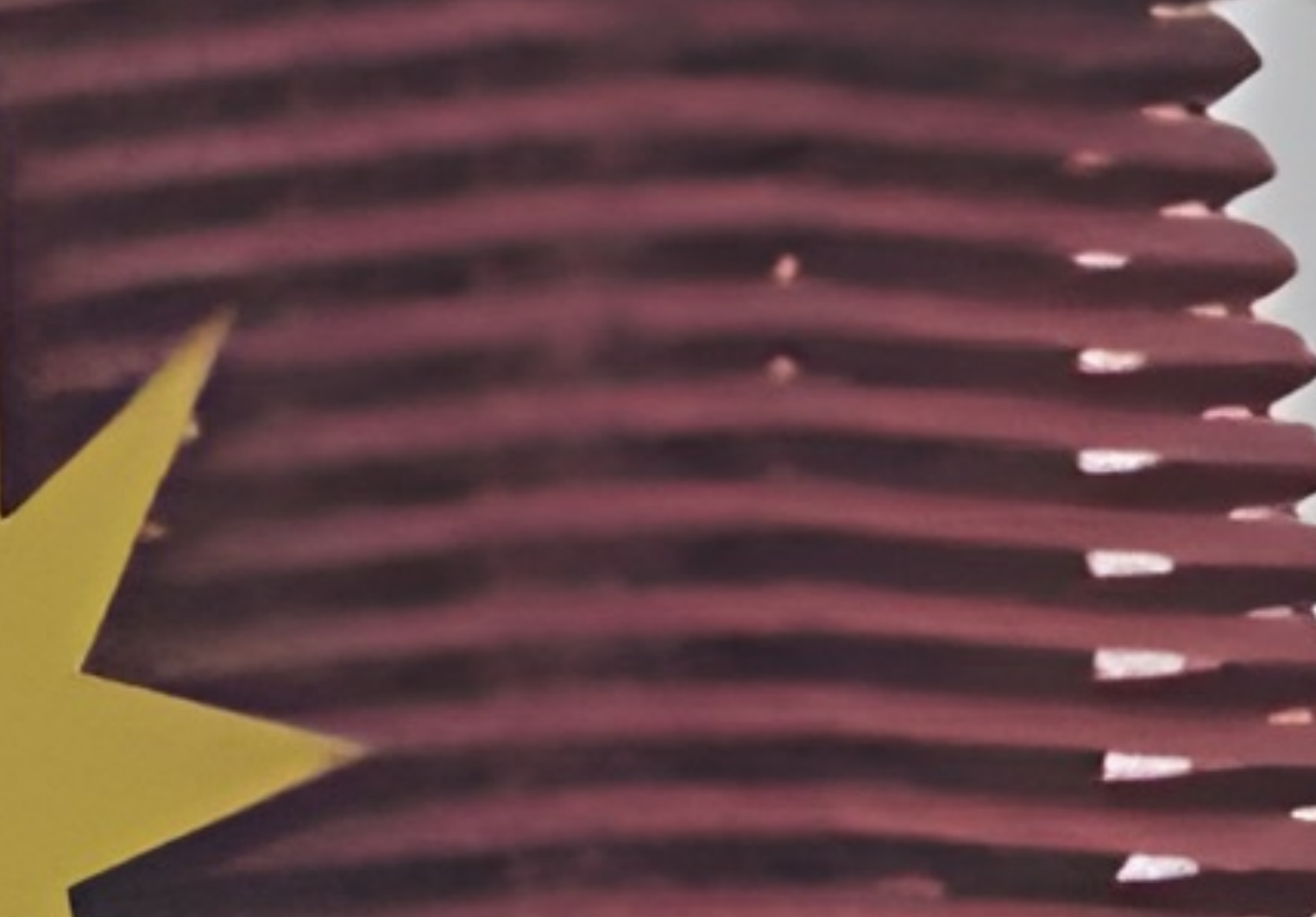}
    \InsertSubfig{0.15}{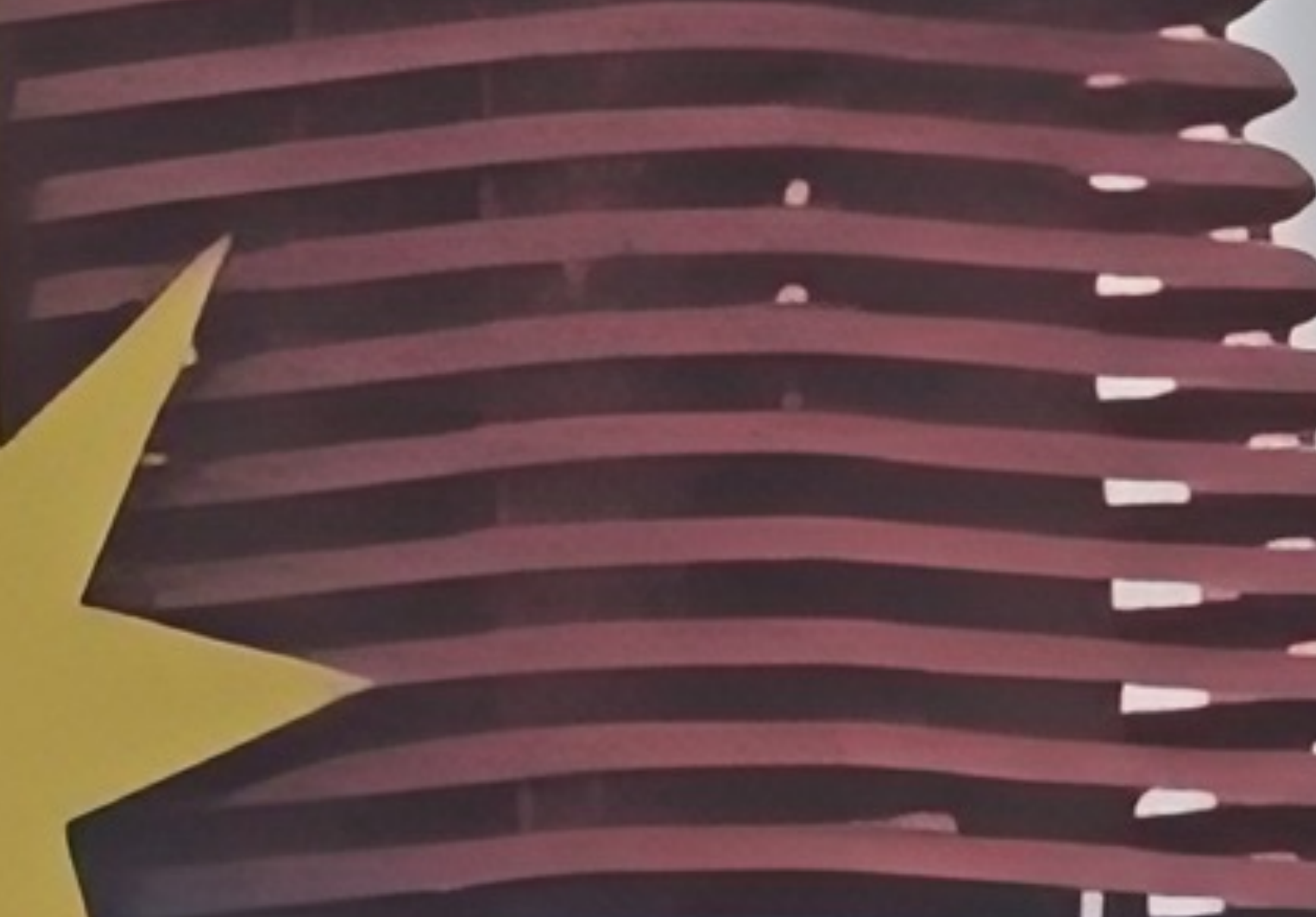}
    \InsertSubfig{0.15}{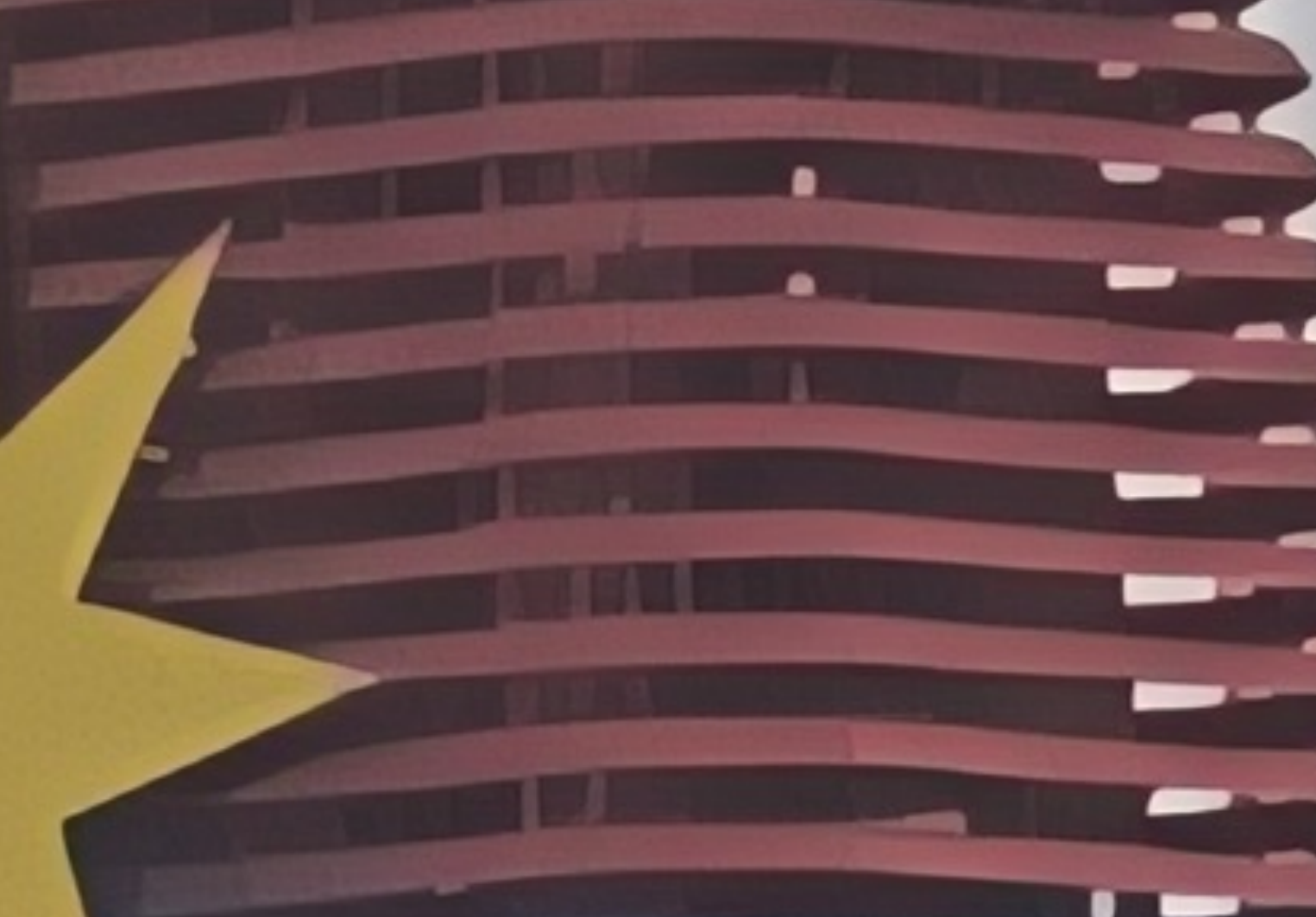}

    \InsertSubfig{0.15}{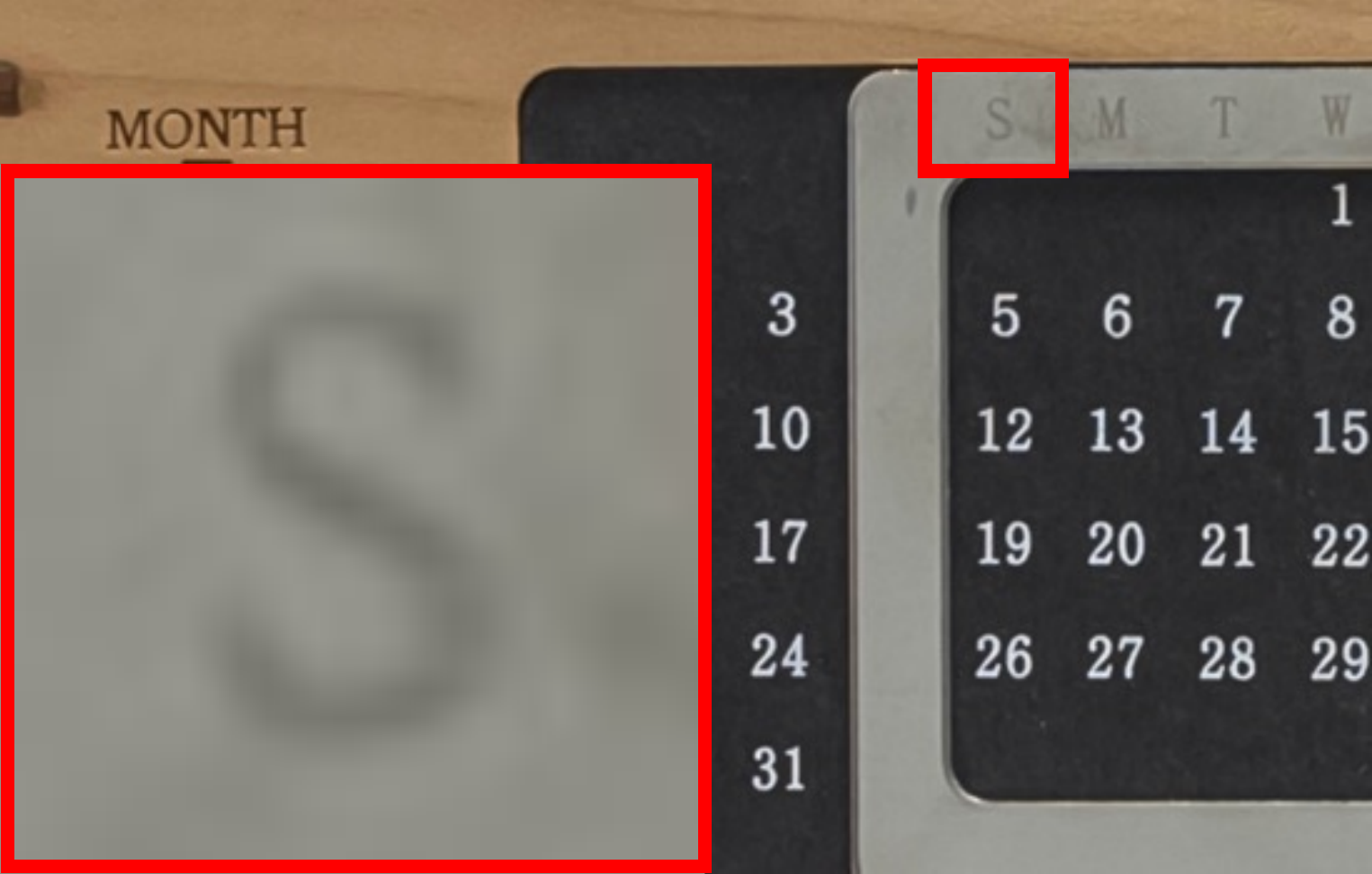}
    \InsertSubfig{0.15}{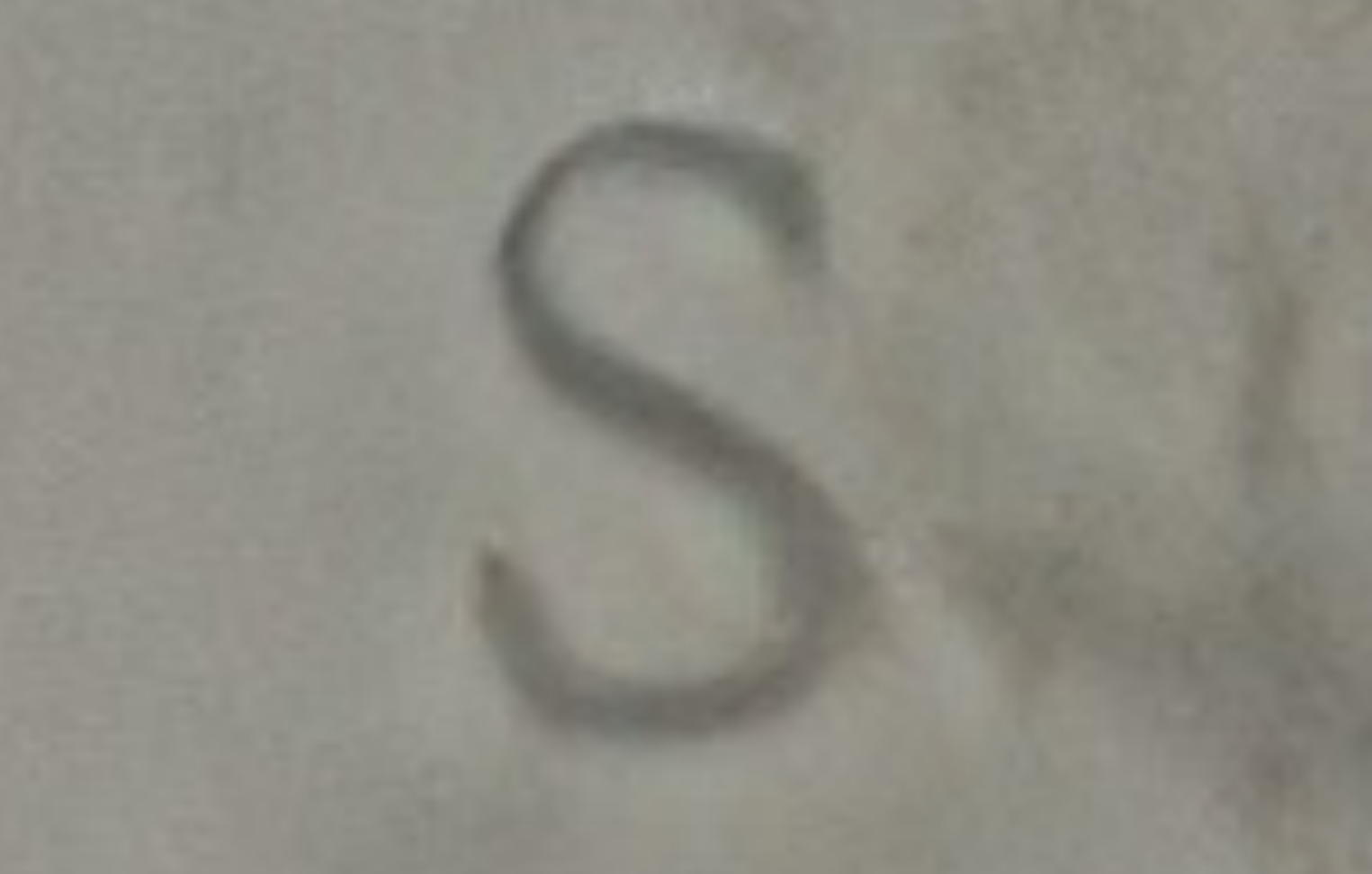}
    \InsertSubfig{0.15}{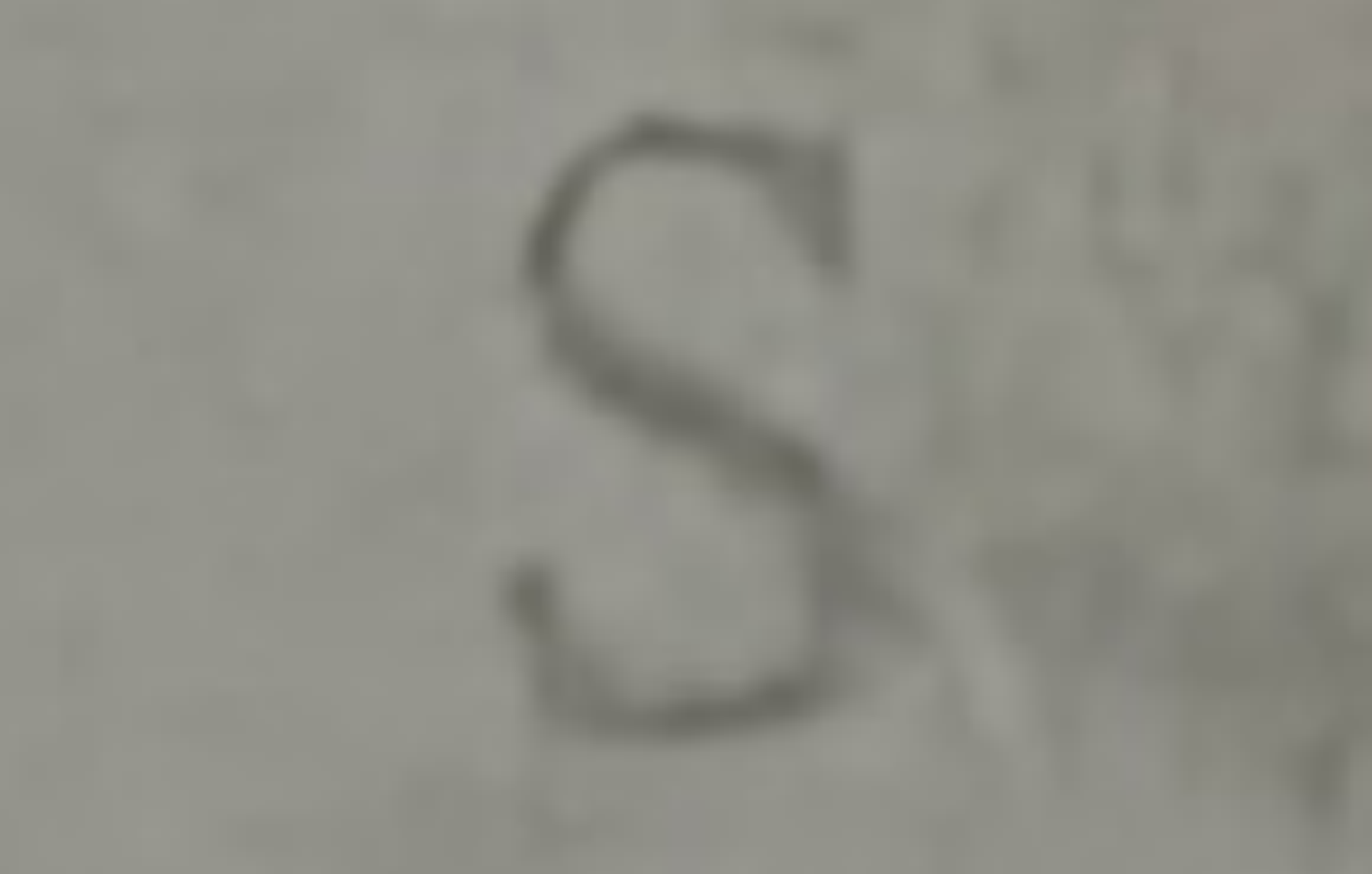}
    \InsertSubfig{0.15}{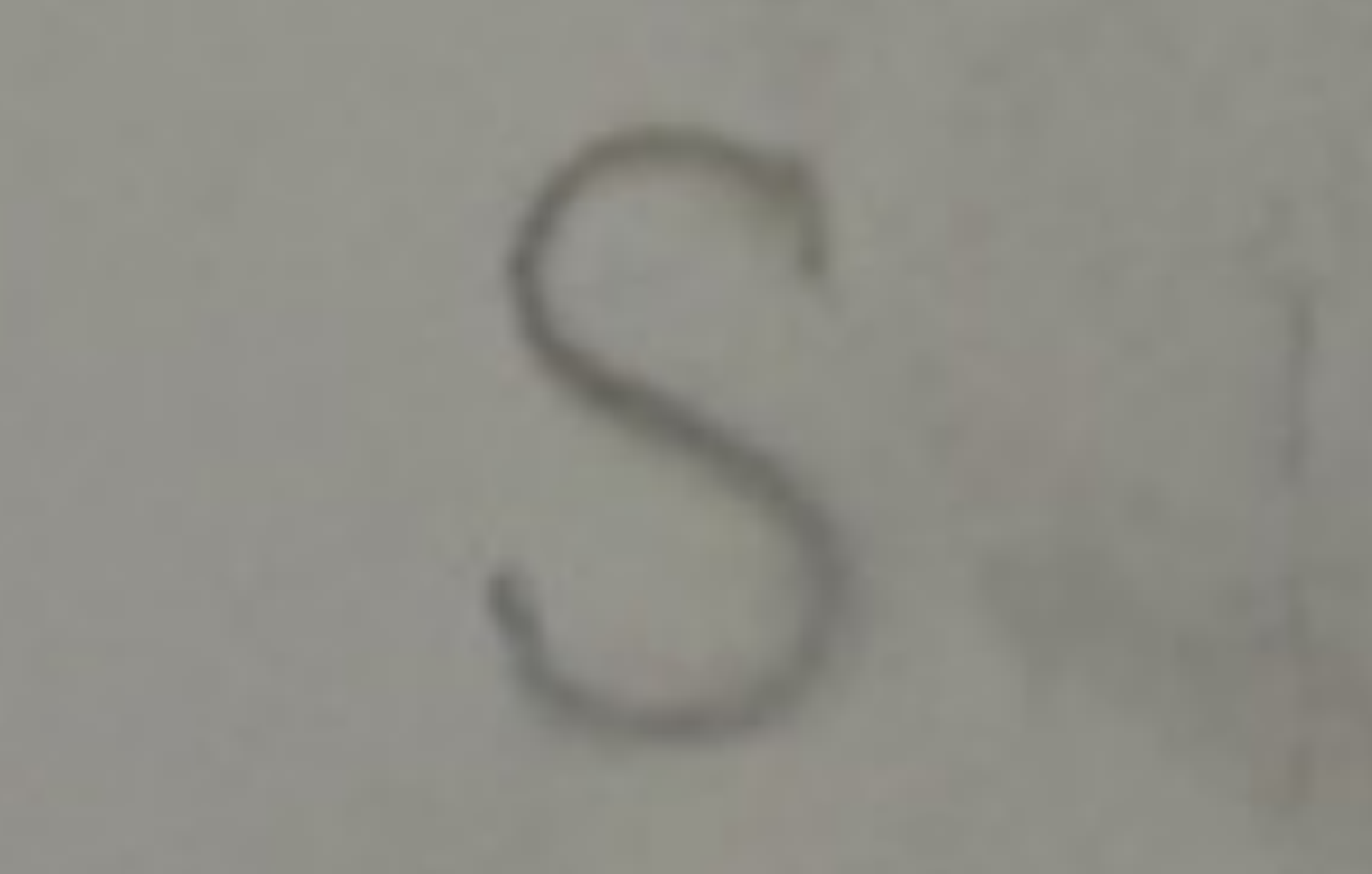}
    \InsertSubfig{0.15}{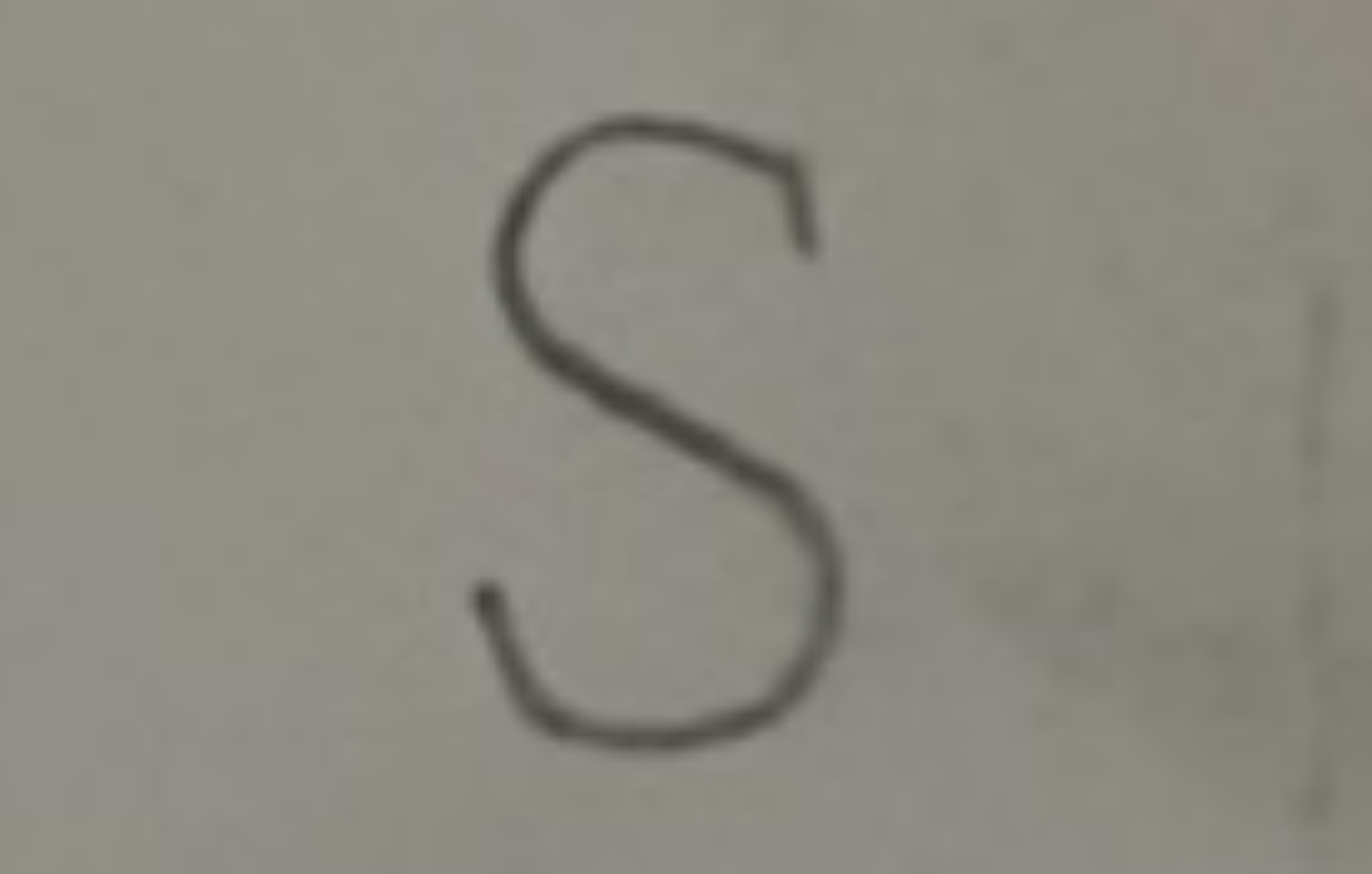}
    \InsertSubfig{0.15}{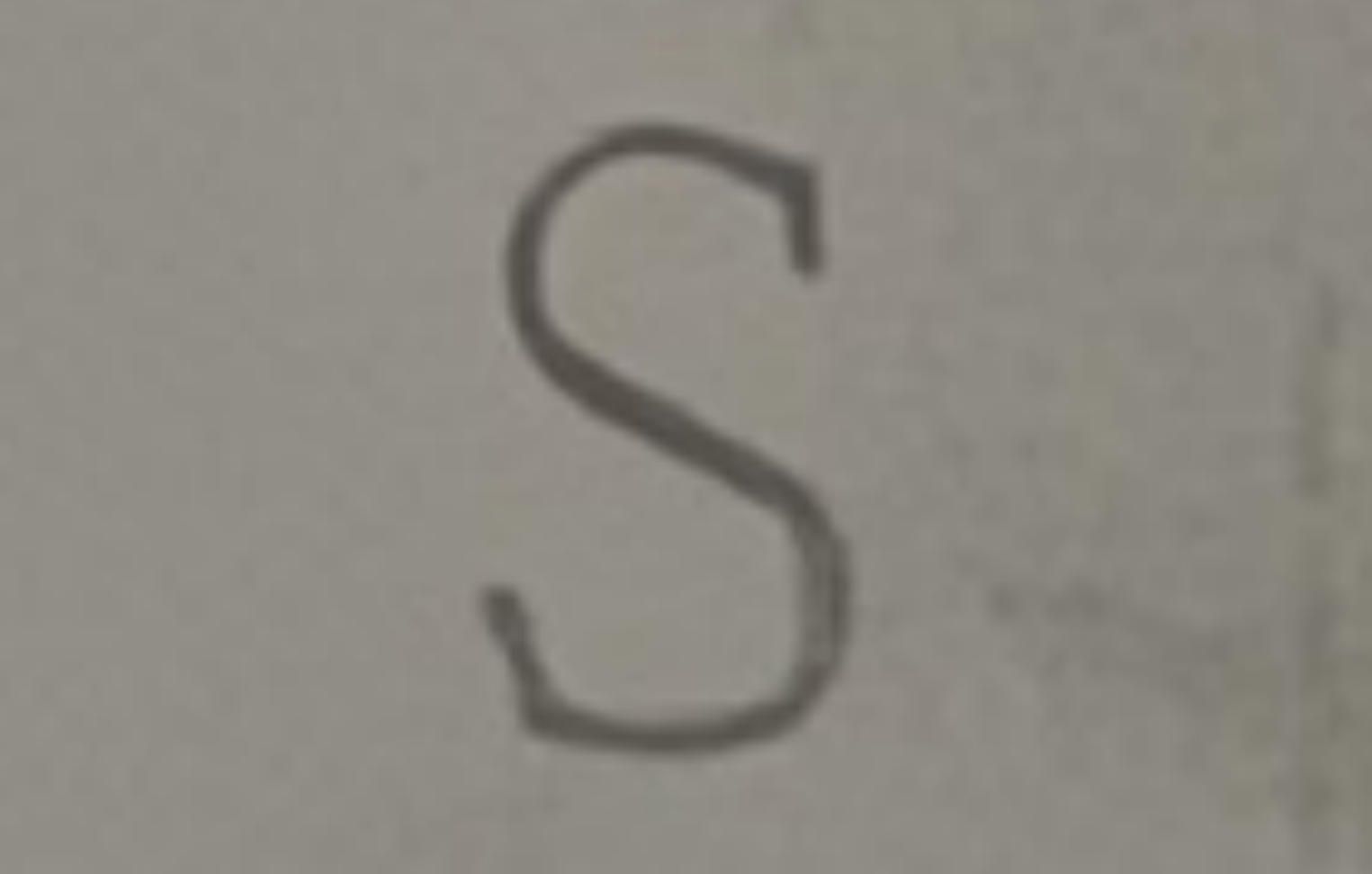}
    
    \InsertSubfigWithCap{0.15}{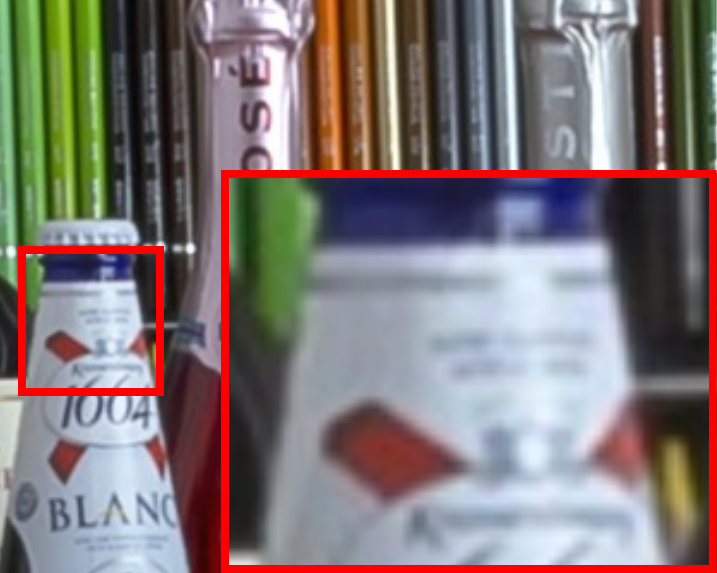}{LR}
    \InsertSubfigWithCap{0.15}{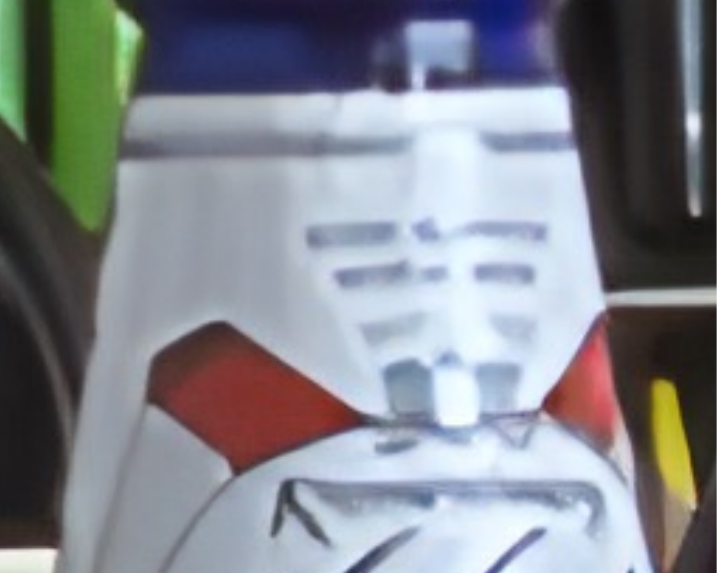}{FeMaSR}
    \InsertSubfigWithCap{0.15}{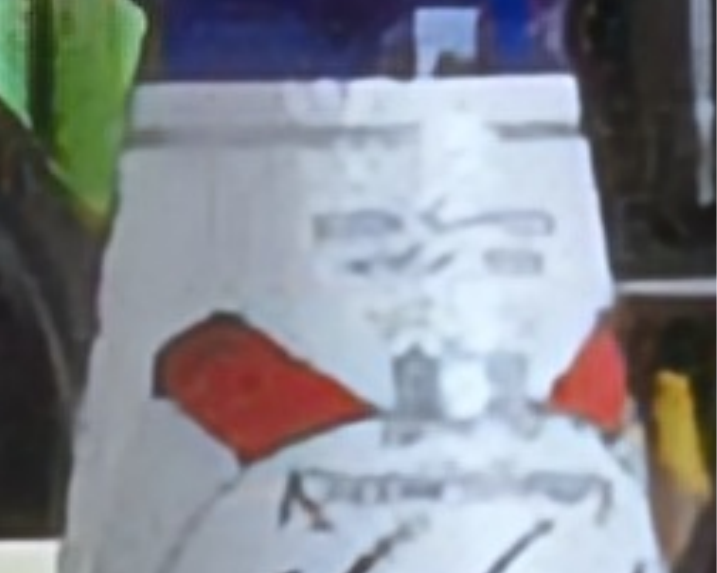}{DASR}
    \InsertSubfigWithCap{0.15}{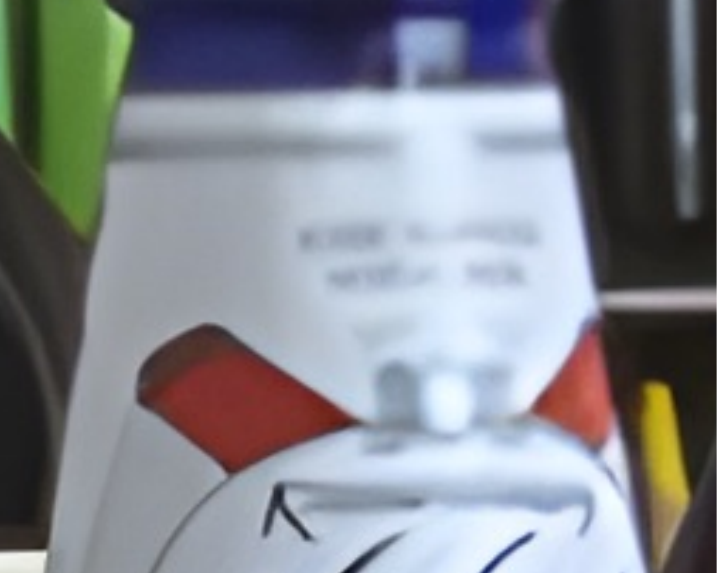}{RE}
    \InsertSubfigWithCap{0.15}{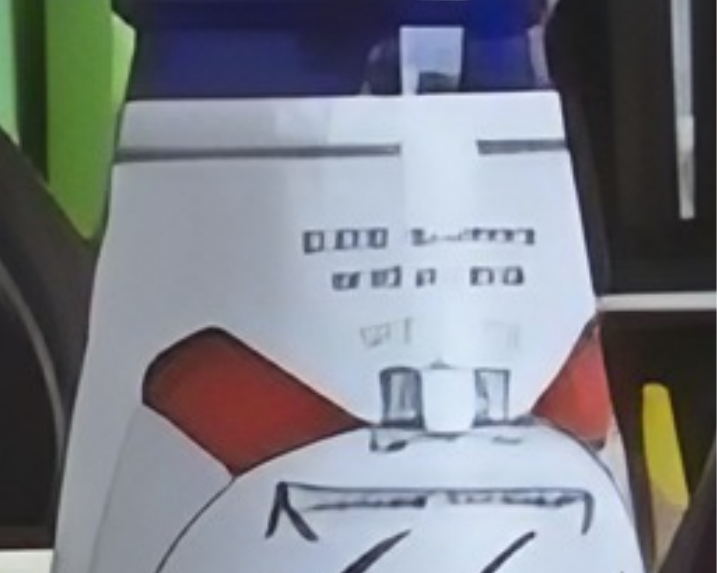}{RE\textcolor{red}{\textbf{+ours}}}
    \InsertSubfigWithCap{0.15}{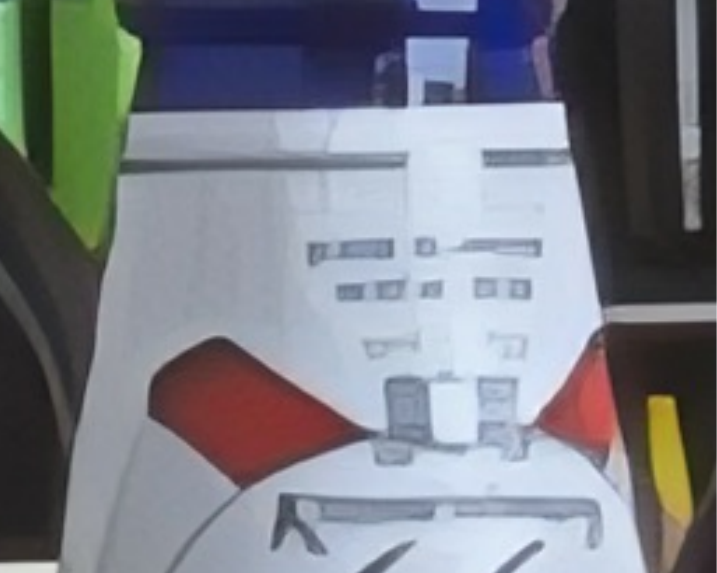}{BG\textcolor{red}{\textbf{+ours}}}
    \caption{Qualitative comparisons. Inputs are from the RealSR-Canon dataset. Columns (e) and (f) are the results of applying our methods on RealESRGAN (RE) and BSRGAN (BG). FeMaSR and DASR can not predict reasonable details.}
    \label{fig:qualitative_cmp}
\end{figure}

%% file: tables/abla_collection.tex
\begin{table}[!h]
    \centering
    \caption{Collection of ablations on EMA and choice of measurement in \cref{eq:ct}. \textbf{Bold} and \underline{underline} mark the best and second best scores. SingleFixed uses the same pre-pretrained model to initialize $M_S$ and $M_G$ as the EMA version but fixes the weights of $M_G$ during training, resulting in inferior performance. Substituting CE in \cref{eq:ct} to $\ell_1$ biases the fidelity scores. }
    \resizebox{0.65\linewidth}{!}{
    \begin{tabular}{l|c c |c c|cc}
    \hline
    \hline
     \multirow{2}{*}{Method} & \multicolumn{2}{c|}{Canon} & \multicolumn{2}{c|}{NTIRE20} & \multicolumn{2}{c}{Olympus} \\
     \cline{2-7}
     & PSNR & NRQM& PSNR & NRQM& PSNR & NRQM\\
     \hline
     Generalist &
    {24.74} & \Best{6.0649} &{25.08} &{6.1213} &26.73 &{5.2530}\\
    \hline
     SingleFixed& {25.71}&{5.9803} &{25.75}&{6.0938} & \Second{{27.47}}&{5.2071}\\
     $\ell_1$& \Best{{26.17}}& {5.8061} & \Best{{26.44}}&\Second{{6.1997}} & \Best{{27.72}}&{4.9602}\\
    \hline
    Ours &
    \Second{25.94} & \Second{6.0647} &\Second{26.42} &\Best{6.2263} &27.40 &\Best{5.3722}\\
    \hline
    \end{tabular}}
    \label{tab:abla_collection}
\end{table}